\pdfoutput=1    

\documentclass[twocolumn,floatfix,showpacs,preprintnumbers,amsmath,amssymb,a4paper,pra]{revtex4}                                          

\usepackage{graphicx}    
\usepackage{dcolumn}     
\usepackage{bm}          
\usepackage[tight,TABTOPCAP,FIGTOPCAP]{subfigure}   
\usepackage{xcolor}

\graphicspath{{Figures/},.}

\newcounter{tr}
\newcommand{\ldb}{[\![}
\newcommand{\rdb}{]\!]}
\newcommand{\bgamma}{\mbox{\boldmath$\gamma$} }
\newcommand{\bn}{{\bf n}}
\newcommand{\sctr}[1]{\begin{subequations}tcounter{tr}{#1}}

\newcommand{\w}{\omega}

\newcommand{\bx}{{\bf x}}
\newcommand{\bq}{{\bf q}}

\newcommand{\kp}{k_{\|}}
\newcommand{\xp}{x_{\|}}
\newcommand{\qp}{q_{\|}}

\newcommand{\aokp}{\alpha_0(k_{\|})}
\newcommand{\aoqp}{\alpha_0(q_{\|})}

\newcommand{\zxp}{\zeta ({\bf x}_{\|})}
\newcommand{\bkp}{{\bf k}_{\|}}

\newcommand{\la}{\langle}
\newcommand{\ra}{\rangle}

\newcommand{\bxp}{{\bf x}_{\|}}
\newcommand{\bqp}{{\bf q}_{\|}}

\newcommand{\sfr}{^{\frac{1}{2}}}

\newcommand{\p}{\partial}

\newcommand{\nn}{\nonumber}

\newcommand{\mbold}[1]{\mbox{\boldmath${#1}$}}

\begin{document}


\title{The Scattering of Electromagnetic Waves from Two-Dimensional
  Randomly Rough Perfectly Conducting Surfaces: The Full Angular
  Intensity Distribution}

\author{Ingve Simonsen} 
\email{Ingve.Simonsen@phys.ntnu.no}
\homepage{http://web.phys.ntnu.no/~ingves} 

\affiliation{Department of
  Physics, Norwegian University of Science and Technology (NTNU),
  NO-7491 Trondheim, Norway}

\author{Alexei A. Maradudin} 
\email{aamaradu@uci.edu}

\author{Tamara A. Leskova} 
\email{tleskova@uci.edu}

\affiliation{Department of Physics and Astronomy and Institute for
  Surface and Interface Science, University of California, Irvine CA
  92697, U.S.A.}

\date{\today}

\begin{abstract}
  By a computer simulation approach we study the scattering of $p$- or
  $s$-polarized light from a two-dimensional, randomly rough,
  perfectly conducting surface. The pair of coupled inhomogeneous
  integral equations for two independent tangential components of the
  magnetic field on the surface are converted into matrix equations by
  the method of moments, which are then solved by the biconjugate
  gradient stabilized method. The solutions are used to calculate the
  mean differential reflection coefficient for given angles of
  incidence and specified polarizations of the incident and scattered
  fields. The full angular distribution of the intensity of the
  scattered light is obtained for strongly randomly rough surfaces by
  a rigorous computer simulation approach.
\end{abstract}

\pacs{42.25.-p; 41.20.-q}

\keywords{scattering, two-dimensional randomly rough surfaces,
  perfect conductor, rigorous computer simulations}

\maketitle


\section{Introduction}

Theoretical/computational studies of the scattering of light from
two-dimensional randomly rough perfectly conducting surfaces are
carried out primarily for two reasons. These are that a perfectly
conducting surface is a good approximation to a finitely conducting
surface in the far infrared region of the optical spectrum, but
computationally less intensive to study than a finitely conducting
surface, and that the development of computational methods for
calculations of scattering from rough perfectly conducting surfaces
can serve as the first step in the development of methods that can be
used in calculations of scattering from rough finitely conducting
surfaces.

In the earliest numerical studies of the scattering of light from a
two-dimensional randomly rough perfectly conducting surface~\cite{1},
the pair of coupled inhomogeneous integral equations for two
independent tangential components of the total magnetic field on the
rough surface obtained from scattering theory was first converted into
a pair of coupled inhomogeneous matrix equations by the methods of
moments~\cite{2}. The system of matrix equations was then solved by
Neumann-Liouville iteration. This is a formally exact approach, but
one that is computationally intensive. It is an $O(MN^2)$ approach,
where $N$ is the number of unknowns to be determined and $M$ is the
number of iterations

Subsequent work on this problem has proceeded in two directions. One
is the exact solution of the integral equations of scattering theory
by numerical methods that are faster than a straightforward
application of the method of moments followed by an iterative solution
of the resulting matrix equation. For example, Wagner {\it et
  al.}~\cite{3} have developed a fast multipole Fast Fourier Transform
method to calculate the scattering of an electromagnetic wave from a
small height two-dimensional randomly rough perfectly conducting
surface that is an $O(N\ln N)$ method. For rougher two-dimensional
perfectly conducting surfaces they have shown that the multi-level
fast multipole algorithm, also an $O(N\ln N)$ method, is more
efficient.

The other direction that has been taken is the approximate solution of
the exact integral equations. In the sparse-matrix flat-surface
iterative approach of Tsang {\it et al.}~\cite{4,5}, the matrix
elements connecting two close points on the surface are treated
exactly, while those connecting two distant points are treated
approximately, in an iterative solution of the matrix equations
obtained by the method of moments. This approach has been applied to
the study of the scattering of electromagnetic waves from a
two-dimensional randomly rough perfectly conducting
surface~\cite{6,7}. It has been elaborated and made faster by Johnson
and his colleagues, resulting in an $O(N)$ method in some cases, and
has been applied to the scattering of electromagnetic waves from a
two-dimensional randomly rough perfectly conducting
surface~\cite{8}. Soriano and Saillard~\cite{9} have developed a
sparse-matrix flat-surface iterative approach, in which the matrix
equations are solved by an iterative Krylov method, the biconjugate
gradient stabilized method~\cite{10}.

In this paper we return to the approach used in~\cite{1}, where the
sparse-matrix flat-surface approximation is not used: the matrix
elements connecting two points are calculated accurately for all
separations of the two points. However, the resulting matrix equations
are solved here by the biconjugate gradient stabilized method instead
of by Neumann-Liouville iteration, as in~\cite{1}. We show that this
approach, together with the increase in computational power
since~\cite{1} was written, provides a simple and reliable way of
calculating the mean differential reflection coefficient for given
angles of incidence and specified polarizations of the incident and
scattered fields, with a modest expenditure of CPU time.

\smallskip

This paper is organized as follows: We start by presenting the
scattering geometry considered (Sec.~\ref{Sec:ScattGeometry}) followed
by the mathematical formulation of the scattering problem
(Sec.~\ref{Sec:Formulation}), including the central integral equation
on which the computer simulations are based. Section~\ref{Sec:Results}
is devoted to the presentation and discussion of the numerical results
obtained from a rigorous computer simulation approach based on an
integral equation for the surface currents derived in
Sec.~\ref{Sec:Formulation}. A detailed discussion of the numerical
aspects of such calculations is given in
Sec.~\ref{Sec:Numerics}. Finally the conclusions that can be drawn
from this work are presented in Sec.~\ref{Sec:Conclusions}.

\section{Scattering Geometry}
\label{Sec:ScattGeometry}

\begin{figure}[th]
  \begin{center}
    \includegraphics*[width=0.95\columnwidth]{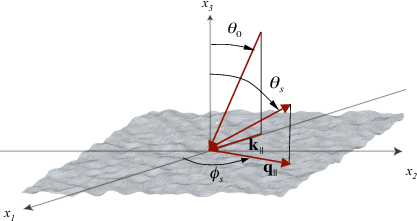}
  \end{center}
  \caption{(Color online) A sketch of the scattering geometry considered in the
    present work, where the coordinate system used and angles of
    incidence and scattering are defined.}
  \label{geom}
\end{figure}

The physical system we consider in this work consists of vacuum in the
region $x_3>\zxp$, where $\bxp=(x_1,x_2,0)$, and a perfect conductor
in the region $x_3<\zxp$ [Fig.~\ref{geom}]. The surface profile
function $\zxp $ is assumed to be a single-valued function of $\bxp $
that is differentiable with respect to $x_1$ and $x_2$, and
constitutes a stationary, zero-mean, isotropic, Gaussian random
process defined by $\la \zxp \zeta(\bxp')\ra=\delta^2
W(|\bxp-\bxp'|)$, where the angle brackets denote an average over the
ensemble of realizations of the surface profile function, and
$\delta=\la\zeta^2(\bxp)\ra\sfr $ is the rms height of the surface. In
the numerical calculations carried out in the present work we will
assume a Gaussian form for $W(|\bxp - \bxp '|)$, namely $W(|\bxp -
\bxp '|) = \exp [-(\bxp - \bxp ')^2/a^2]$, where $a$ is the transverse
correlation length of the surface roughness. Each realization of the
surface profile function with these  properties is generated
numerically by a two-dimensional version of the filtering method used
in~\cite{11}.

\section{Formulation}
\label{Sec:Formulation}


\subsection{Integral Equation}

The starting point for our analysis is the Stratton-Chu
formula~\cite{12} for the magnetic field in the vacuum,
\begin{align}
  &\theta(x_3-\zxp ) {\bf H}^>(\bx |\w ) = 
  {\bf H}(\bx |\w )_{inc} \nonumber \\
  & \qquad  + \frac{1}{4\pi} \int d^2\xp '\left[ \nabla' g_0 (\bx |\bx ')\right]_{x'_3=\zeta (\bxp ')}\times {\bf J}_{H}(\bxp '|\w  ),\label{eq:1}
\end{align}
where $\theta (z)$ is the Heaviside unit step function, and ${\bf
  H}(\bx|\w)_{inc}$ is the magnetic component of the incident
field.

The function $g_0(\bx |\bx ')$ is the scalar free-space Green's
function and has the representations
\begin{subequations}
\begin{eqnarray}
  g_0(\bx |\bx ')
  &=& \frac{\exp\left[i\frac{\w}{c}|\bx-\bx '|\right]}{ |\bx-\bx '| } 
                      \label{g0-x}\\
  &=&  \int\!\frac{d^2\qp}{(2\pi )^2} \; \frac{2\pi i}{\aoqp}
    \exp\left[i \bqp\cdot(\bxp-\bxp')\right]\quad    \nonumber  \\
  &&   \hspace{1.8cm}   \times \exp\left[i\aoqp |x_3-x_3'|\right],\label{g0-q}
\end{eqnarray}
\label{g0}
\end{subequations} where
\begin{align}
 & \aoqp = \sqrt{\left(\frac{\w}{c}\right)^2-\qp^2},
  \;\; Re\,\aoqp>0, Im\,\aoqp>0, \nonumber \\
  &
\end{align}
and $\w$ and $c$ are the frequency and speed of light in vacuum.  In
writing Eq.~(\ref{eq:1}) we have assumed the time dependence
$\exp(-i\w t)$ for the field, but have not indicated this
explicitly. The (electric) surface current density ${\bf J}_H(\bxp |\w
)$ is defined by ${\bf J}_H(\bxp |\w ) = [\bn \times {\bf H}^>(\bx |\w
)]\big|_{x_3 = \zxp}$, where $\bn =(-\zeta_1(\bxp), -\zeta_2(\bxp),1
)$ is a vector that is normal to the surface $x_3=\zxp$ at each point
of it, directed into the vacuum, and we have introduced the notation
$\zeta_j(\bxp)=\p \zxp/\p x_j$ $(j=1,2)$.

On evaluating Eq.~(\ref{eq:1}) at $x_3=\zxp +\eta$ and at $x_3=\zxp
-\eta$, where $\eta$ is a positive infinitesimal, adding the resulting
two equations, and taking the vector cross product of the sum with
$\bn $, we obtain the integral equation satisfied by the surface
current ${\bf J}_{H}(\bxp|\w)$,
\begin{widetext}
\begin{eqnarray} 
  {\bf J}_H(\bxp |\w ) &=& 
      2{\bf J}^{(i)}_{H}(\bxp |\w)
      + \frac{1}{2\pi} P\int \! d^2\xp ' \; {\bf n}\times \left(\ldb \nabla
        ' 
        g_0(\bx |\bx ')\rdb \times {\bf J}_H(\bxp '|\w )\right),
  \label{eq:2}
\end{eqnarray}
\end{widetext}
where ${\bf J}^{(i)}_H(\bxp |\w ) = {\bf n}\times {\bf H}(\bx |\w
)_{inc}\big|_{x_3 = \zxp} $, $P$ denotes the Cauchy principal value,
and we have simplified the notation by introducing the definition
\begin{eqnarray}
  \ldb f(\bx |\bx ')\rdb 
  &=&  f(\bx |\bx ')\bigg|_{\substack{ x_3=\zxp \\ x'_3=\zeta(\bxp ') }}.
   \label{eq:3}
\end{eqnarray}

The system of three equations (\ref{eq:2}) can be reduced to a system
of two equations through the use of the condition ${\bf n}\cdot {\bf
  J}_H(\bxp |\w ) = 0$.  Thus only two components of ${\bf J}_H(\bxp
|\w )$ are independent. We choose ${\bf J}_H(\bxp |\w )_{1,2}$ as the
independent components, while
\begin{eqnarray}
  J_H(\bxp |\w )_3 
  &=& 
   \zeta_1(\bxp )J_H(\bxp |\w )_1 
  +\zeta_2(\bxp )J_H(\bxp |\w )_2 .
  \quad
  \label{eq:4}
\end{eqnarray}

From Eq.~(\ref{eq:2}) we find with the aid of Eq.~(\ref{eq:4}) that
the components ${\bf J}_H(\bxp |\w )_{1,2}$ satisfy the following pair
of equations: 
\begin{widetext} 
\begin{subequations}
\begin{eqnarray}
  J_H(\bxp |\w )_1 &=& 
     2J^{(i)}_H(\bxp |\w )_1
       - \frac{1}{2\pi}P\int \! d^2\xp ' \;\biggl\{ \Bigl[  g_3^{(0)}(\bxp
       |\bxp ') -  g_1^{(0)}(\bxp |\bxp ')\zeta_1(\bxp ')
       -\zeta_2(\bxp ) g_2^{(0)}(\bxp |\bxp ') \Bigr] J_H(\bxp '|\w
       )_1 \nn\\
  && \qquad \qquad \qquad \qquad \qquad \qquad \qquad 
      + g_1^{(0)}(\bxp |\bxp ')\Bigl[ \zeta_2(\bxp ) - \zeta_2(\bxp' )\Bigr]J_H(\bxp '|\w )_2\biggr\}
  \label{eq:5a}
\end{eqnarray}
\begin{eqnarray}
  J_H(\bxp |\w )_2 &=& 
     2J_H^{(i)}(\bxp |\w )_2
       - \frac{1}{2\pi} P\int\!d^2\xp ' \; \biggr\{g_2^{(0)}(\bxp |\bxp ')
          \Bigl[\zeta_1(\bxp ) - \zeta_1(\bxp' )\Bigr]J_H(\bxp '|\w )_1
          \qquad \qquad \qquad \qquad \qquad \qquad \nn\\
   && 
   \qquad \qquad \qquad \qquad \qquad 
   +\Bigl[g_3^{(0)}(\bxp |\bxp ') - g_2^{(0)}(\bxp |\bxp ')\zeta_2(\bxp ') 
      - \zeta_1(\bxp ) g_1^{(0)}(\bxp |\bxp ')\Bigr] J_H(\bxp '|\w )_2 \biggr \},
      \label{eq:5b}
\end{eqnarray}
\label{eq:5}
\end{subequations} 
where 
\begin{eqnarray}
  g^{(0)}_l(\bxp|\bxp ') &=& \ldb \frac{\p}{\p x_l}g_0(\bx|\bx ')\rdb 
   =  (x_l-x'_{l}) \left[\frac{i(\w/c)}{|\bx-\bx '|^2}-\frac{1}{|\bx
       -\bx '|^3}\right]\exp[i(\w/c)|\bx -\bx '|]
         \bigg|_{\substack{ x_3=\zxp \\ x'_3=\zeta(\bxp ') }}. 
  \label{eq:6}
\end{eqnarray}
Equations (\ref{eq:5}) are solved by converting them into a pair of
coupled matrix equations. This is done by generating a realization of
the surface profile function on a grid of $N^2$ points within a square
region of the $x_1x_2$ plane of edge $L$, where the ratio $L/N=\Delta
x$ is chosen to be $\Delta x= \lambda/7$, with $\lambda$ the
wavelength of the incident field. The integrals over this region in
Eqs.~(\ref{eq:5}) are carried out by means of a two-dimensional
version of the extended midpoint method~\cite{13}, and the values of
${\bf J}_H(\bxp |\w )_{1,2}$ are calculated at the points of this
grid. The resulting matrix equations are then solved by means of the
biconjugate gradient stabilized method~\cite{10}. Once ${\bf J}_H(\bxp
|\w )_{1}$ and ${\bf J}_H(\bxp |\w )_{2}$ have been obtained in this
way, ${\bf J}_H(\bxp |\w )_{3}$ is obtained from Eq.~(\ref{eq:4}).

\subsection{Scattered Field}

With the surface current ${\bf J}_H(\bxp |\w )$ in hand, one is ready
to start approaching the calculation of the scattered field. To this
end, let us start by writing the scattered electric field (in the far
zone) in the form
\begin{eqnarray} 
  {\bf E}(\bx|\w)_{sc} 
  &=& 
      \int\!\frac{d^2\qp}{(2\pi)^2} \; 
            \mbold{\mathcal E}({\bf q}_+,\w) \,
            \exp[i\bq_+\cdot\bx], 
             \nonumber \\
  &=&
    \int\!\frac{d^2\qp}{(2\pi)^2} \; 
       \left[
              {\mathcal E}_p({\bf q}_+,\w) \, \hat{\bgamma}_p({\bf q}_+,\w)
            + {\mathcal E}_s({\bf q}_+,\w) \, \hat{\bgamma}_s({\bf q}_+,\w)
       \right]
       \exp[i\bq_+\cdot\bx],
       \label{eq:7-new}
\end{eqnarray} 
\end{widetext}
where 
${\mathcal E}_\nu = \mbold{\mathcal E} \cdot \hat{\bgamma}_\nu$
($\nu=p,s$).
In writing Eq.~(\ref{eq:7-new}) we have introduced the (unit)
polarization vectors $\hat{\bgamma}_\nu({\bf q}_\pm,\w)$ for $p$- and
$s$-polarized scattered light that are mutually orthogonal and also
orthogonal to the wave-vector ${\bf q}_\pm$. They can, in accordance
with Sipe~\cite{Sipe}, be defined as
\begin{subequations}
  \label{eq:pol-vectors}
\begin{eqnarray}
  \label{eq:pol-vectors-p}
  \hat{\bgamma}_s({\bf q}_\pm,\w) 
     &=& 
     \frac{ {\bf q}_\pm \times \hat{\bf x}_3}{ \left| {\bf q}_\pm \times \hat{\bf x}_3\right|}
     \;=\
       \hat{\bf q}_\parallel \times \hat{\bf x}_3,
       \\
  \label{eq:pol-vectors-s}
  \hat{\bgamma}_p({\bf q}_\pm,\w) 
     &=& 
        \hat{\bgamma}_s({\bf q}_\pm,\w) \times \hat{\bf q}_\pm 
        \nonumber \\
     &=&  
          \frac{ 
           q_\parallel\,\hat{\bf x}_3  \mp \alpha_0(q_\parallel,\omega)\, \hat{\bf q}_\parallel 
          }{ 
            \w/c 
          }, 
\end{eqnarray}
where we have introduced the wave-vector for {\em upward} (${\bf
  q}_+$) and {\em downward} (${\bf q}_-$) propagating (plane) waves 
\begin{eqnarray} 
  \label{eq:pol-vectors-q}
  {\bf q}_\pm( {\bf q}_\parallel, \w)   
  &=& 
  {\bf q}_\parallel \pm \alpha_0(q_\parallel) \hat{\bf x}_3. 
\end{eqnarray}  
\end{subequations} 

From Eqs.~(\ref{eq:pol-vectors}) it is readily shown that the set
$\{\hat{\bgamma}_p({\bf q}_\pm,\w)$, $\hat{\bgamma}_s({\bf
  q}_\pm,\w)$, $\hat{\bf q}_\pm({\bf q}_\parallel,\w)\}$ forms a
(right-handed) orthonormal triad.  This implies, for instance,
suppressing the function arguments for simplicity, that
$\hat{\bgamma}_\mu \cdot \hat{\bgamma}_{\nu} = \delta_{\mu\nu}$,
${\bf q}_\pm \cdot \hat{\bgamma}_\nu=0$ as well as
\begin{subequations}
  \label{eq:pol-vectors-relations}
\begin{eqnarray} 
  \hat{\bgamma}_s &=& \hat{\bf q}_\pm  \times \hat{\bgamma}_p,
    \label{eq:pol-vectors-relations-B}
  \\
  \hat{\bgamma}_p &=& - \hat{\bf q}_\pm \times \hat{\bgamma}_s, 
  \label{eq:pol-vectors-relations-C}  
  \\
  \hat{\bf q}_\pm  &=& \hat{\bgamma}_p \times \hat{\bgamma}_s. 
  \label{eq:pol-vectors-relations-A}
\end{eqnarray}  
\end{subequations}

With the use of one of the Maxwell's equations (Faraday's law),
$\nabla \times {\bf E}= i(\w/c) {\bf H}$, and
Eqs.~(\ref{eq:pol-vectors-relations}), it follows from
Eq.~(\ref{eq:7-new}) that the scattered magnetic field can be written
\begin{widetext}
\begin{eqnarray} 
  {\bf H}(\bx|\w)_{sc} &=& 
       \int\!\frac{d^2\qp}{(2\pi)^2} \; 
       \left[
              {\mathcal E}_p({\bf q}_+,\w) \, \hat{\bgamma}_s({\bf q}_+,\w) 
            - {\mathcal E}_s({\bf q}_+,\w) \, \hat{\bgamma}_p({\bf q}_+,\w) 
       \right]
       \exp[i\bq_+\cdot\bx].
       \label{eq:7}
\end{eqnarray} 
On the other hand, the scattered magnetic field is also given in terms
of the surface current ${\bf J}_H(\bxp |\w )$ by the second term on
the right-hand side of Eq.~(\ref{eq:1}), and with the use of
Eq.~(\ref{g0-q}) one is led to~($\nu=p,s$)
\begin{eqnarray}
  {\mathcal E}_\nu({\bf q}_+ ,\w ) &=& 
     -\frac{(\w /c)}{2\aoqp} 
       \int \! d^2\xp \;  
       \hat{\bgamma}_\nu ({\bf q}_+ ,\w ) \cdot {\bf J}_H(\bxp |\w )\;
       \exp[-i\bq_+\cdot\bx].
    \label{eq:8} 
\end{eqnarray}

The total time-averaged scattered flux is given by the real part of
the $3$-component of the (complex) Poynting vector (${\bf S}^c =
c/(8\pi)\,{\bf E} \times {\bf H}^*$) of the scattered field,
integrated over the plane $x_3=0$. From the fields in the form of
Eqs.~(\ref{eq:7-new}) and (\ref{eq:7}) and the use of
Eqs.~(\ref{eq:pol-vectors-relations}) we find that it is given by
\begin{eqnarray}
  P_{sc} &=& \frac{c^2}{8\pi\w}\int\limits_{\qp<\frac{\w}{c}} \!
  \frac{d^2\qp}{(2\pi )^2} \; 
  \aoqp\left[ 
         \left|{\mathcal E}_p({\bf q}_+,\w)\right|^2
       + \left|{\mathcal E}_s({\bf q}_+,\w)\right|^2
       \right],
  \label{Psc}
\end{eqnarray}
and we recall that ${\bf q}_+={\bf q}_+(\bqp,\w)$, defined in
Eq.~(\ref{eq:pol-vectors-q}), depends on the parallel momentum
$\bqp$. Moreover, the vector $\bqp$ is given in terms of the polar and
azimuthal scattering angles $\theta_s$ and $\phi_s$ by
\begin{eqnarray}
  \bqp &=& \frac{\w}{c}\sin\theta_s\left(\cos\phi_s,\sin\phi_s,0\right).
\end{eqnarray}
The expression given by Eq.~(\ref{Psc}) can then be rewritten as
\begin{eqnarray}
  P_{sc} &=& 
   \frac{c^2}{8\pi\w}\frac{1}{4\pi^2}\left(\frac{\w}{c}\right)^3 
   \int\!d\Omega_s \; \cos^2\theta_s 
  \left[   \left|{\mathcal E}_p({\bf q}_+,\w)\right|^2
         + \left|{\mathcal E}_s({\bf q}_+,\w)\right|^2 
  \right],
  \label{Psc-O}
\end{eqnarray}
\end{widetext}
where $d\Omega_s=\sin\theta_s d\theta_s d\phi_s$ is the element of
solid angle about the scattering direction $(\theta_s,\phi_s)$.

\subsection{Incident Field}

The incident electric field vector that will be considered in this
study, has the form of a (Gaussian) beam propagating in the direction
of 
\begin{align}
  {\bf k} &= \frac{\omega}{c} 
  \left( 
    \sin\theta_0 \cos\phi_0, \, 
    \sin\theta_0 \sin\phi_0, \,
    - \cos\theta_0
  \right),
\end{align}
and is represented by a superposition of incoming
plane waves
\begin{subequations}
  \label{eq:beam}
\begin{eqnarray} 
 {\bf E}(\bx |\w )_{inc} &=& 
  \int\limits_{\qp < \frac{\w}{c}} \! d^2\qp \; 
  \mbox{\boldmath$\mathcal E$}^{(i)}({\bf q}_-,\w )\,
  \exp[i\bq_-\cdot\bx]
  \nonumber \\
  && \hspace{2cm} \times
   \;   W(\bqp,\bkp),
  \label{eq:10-a}
\end{eqnarray}
where $W(\bqp,\bkp)$ denotes an envelope (or window) function, here
defined as 
\begin{eqnarray} 
  W(\bqp,\bkp) &=& \frac{w^2}{2\pi} \exp \left[ - \frac{w^2}{2}(\bqp -\bkp )^2\right] ,
  \label{eq:10-b}
\end{eqnarray}
\end{subequations}
with $w$ its (and the beam's) half width. Note that in the limit of large beam
widths ($w\rightarrow\infty$), the envelope $W(\bqp,\bkp)$ tends towards
$\delta(\bqp-\bkp)$ so that, in this limit, the incident beam becomes
a plane wave.

A beam as defined by Eqs.~(\ref{eq:beam}) does not adhere to the usual
definition of $p$- or $s$-polarized waves since the plane of incidence
is not well-defined in this case (except when $w=\infty$). However, we
will still refer to an incident beam of the form given by
Eqs.~(\ref{eq:beam}) as $p$-polarized if its electric field vector is
in the plane ``of incidence'' defined by the vectors ${\bf k}$ and
$\hat{\bf x}_3$. Therefore, for a $p$-polarized beam, the projection
of its amplitude vector $\mbold{\mathcal E}^{(i)}_p({\bf q}_- ,\w )$
onto the $x_1x_2$-plane will be parallel to ${\bf k}_\parallel$.
Moreover, the vector amplitude for an $s$-polarized beam,
$\mbold{\mathcal E}^{(i)}_s({\bf q}_- ,\w )$, is defined as
\begin{eqnarray}
  \label{eq:Spol_beam}
  \hat{\mbold{\mathcal E}}{}^{(i)}_s({\bf q}_- ,\w ) 
  &=& \
  \hat{\bf q}_-
  \times  
  \hat{\mbold{\mathcal E}}{}^{(i)}_p({\bf q}_- ,\w ),
\end{eqnarray}
similarly to the relation satisfied by the plane-wave polarization
vectors $\hat{\bgamma}_\nu$ ({\it cf.}
Eq.~(\ref{eq:pol-vectors-relations-B})).

Since in this work we are concerned exclusively with isotropic
surfaces, we will, with no loss of generality, assume that the vector
$\bkp$, if non-zero, is parallel to the $x_1$ axis, {\it i.e.}
$\bkp=\kp \hat{\bf x}_1$. Under this assumption the amplitude vector
for a $p$-polarized incident beam, $\mbold{\mathcal E}{}^{(i)}_p({\bf
  q}_- ,\w )$, will lie in the $x_1x_3$-plane, {\it i.e.} its second
component will be zero, which with the condition $\nabla\cdot{\bf
  E}=0$ (or equivalently ${\bf q}_-\cdot
\mbold{\mathcal E}{}^{(i)}({\bf q}_- ,\w )=0$) leads us to define
\begin{subequations}
  \label{eq:11}
  \begin{eqnarray}
\hat{\mbold{\mathcal E}}{}^{(i)}_p({\bf q}_- ,\w ) &=& 
  \frac{ \aoqp\,\hat{\bf x}_1 + q_1\,\hat{\bf x}_3 }{[q^2_1+\alpha^2_0(\qp )]\sfr}.
  \label{eq:11b}
\end{eqnarray}
The amplitude for the corresponding $s$-polarized beam follows from
Eq.~(\ref{eq:Spol_beam}), and, 
with the
use of Eq.~(\ref{eq:pol-vectors-q}), it can be written as
\begin{eqnarray}
  \hat{\mbold{\mathcal E}}{}^{(i)}_s({\bf q}_- ,\w )
  &=& \frac{
    q_1q_2\, \hat{\bf x}_1  -[q^2_1+\alpha^2_0(\qp )]\,\hat{\bf x}_2
    -q_2\aoqp\,\hat{\bf x}_3 
  }{ (\w/c)  \,   
    [q^2_1+\alpha^2_0(\qp )]\sfr}.\nonumber \\
 \label{eq:11a}
\end{eqnarray}
\end{subequations} 
With the beam amplitudes in the form of Eqs.~(\ref{eq:11}) it is
readily established that similar relations to those satisfied by the
plane-wave polarization vectors ({\it e.g.}
Eqs.~(\ref{eq:pol-vectors}) and (\ref{eq:pol-vectors-relations})),
also hold for the polarization amplitudes, $\hat{\mbold{\mathcal
    E}}{}^{(i)}_\nu$, of the Gaussian beam.

Moreover, also note that in the limit of a large beam width
($w\rightarrow \infty$) Eqs.~(\ref{eq:11}) reduce to the plane wave
polarization vectors given previously in Eqs.~(\ref{eq:pol-vectors})
since in this limit ${\bf q}_\parallel={\bf k}_\parallel$ with
$k_\parallel=k_1$. This is another reason for associating the vector
amplitudes of Eqs.~(\ref{eq:11}) with $p$- and $s$-polarized components,
respectively.

With the polarization vectors available for the incident $p$- and
$s$-polarized components of the incident beam, the incident electric
field, of given polarization $\nu$, can according to
Eqs.~(\ref{eq:beam}) and Eqs.~(\ref{eq:11}), be written (assuming unit
amplitude for simplicity) in the following form
\begin{eqnarray} 
{\bf E}_\nu(\bx |\w )_{inc} &=& 
  \int\limits_{q_1 < \frac{\w}{c}}\! d^2\qp \;
  \hat{\mbold{\mathcal E}}{}^{(i)}_\nu({\bf q}_- ,\w )
  \exp [i\bq_-\cdot \bx  ]
  \nonumber \\
  && \hspace{2cm} \times
  \;  W(\bqp,\bkp).
  \label{eq:14}
\end{eqnarray}
In precisely the same way as Eq.~(\ref{eq:7}) was established for the
scattered field, it follows from Eqs.~(\ref{eq:14}) by using
Eqs.~(\ref{eq:Spol_beam}) and relations for $\hat{\mbold{\mathcal
    E}}{}^{(i)}_\nu$ similar to those of
Eqs.~(\ref{eq:pol-vectors-relations}), that the magnetic component of
the incident beam then takes the form
\begin{subequations}
  \label{eq:12}
\begin{eqnarray} 
   {\bf H}_p(\bx |\w )_{inc} &=& 
  \int\limits_{\qp < \frac{\w}{c}} \! d^2\qp \;
  \hat{\mbold{\mathcal E}}{}^{(i)}_s({\bf q}_- ,\w )
  \exp [i\bq_-\cdot \bx  ]
  \nonumber \\
  && \hspace{2cm} \times 
  \; W(\bqp,\bkp),
  \label{eq:12-A}
\end{eqnarray}
for a $p$-polarized beam, and 
\begin{eqnarray}
  {\bf H}_s(\bx |\w )_{inc} &=& - 
  \int\limits_{\qp < \frac{w}{c}} \! d^2\qp \;
  \hat{\mbold{\mathcal E}}{}^{(i)}_p({\bf q}_- ,\w )
  \exp [i\bq_-\cdot \bx  ]
  \nonumber \\
  && \hspace{2cm} \times 
 \; W(\bqp,\bkp), 
  \label{eq:12-B}
\end{eqnarray}
\end{subequations}
for an $s$-polarized beam.

With the incident field in the form of Eqs.~(\ref{eq:14}) and
(\ref{eq:12}), the magnitude of the total time-averaged incident flux
is the same for light of both polarizations, and is given by
\begin{eqnarray}
  P^{(p,s )}_{inc} &=& \frac{c^2}{8\pi\w} p_{inc}, \label{eq:16}
\end{eqnarray}
where 
\begin{widetext}
\begin{subequations}
  \begin{eqnarray}
    p_{inc} & =& w^4 \int\limits_{\qp < \frac{\w}{c}}\! d^2\qp \;
    \aoqp \exp [-w^2 (\bqp - \bkp )^2] \label{eq:17a}\\
        &=&2\pi w^4 \left(\frac{\w}{c}\right)^3\exp\left(-w^2\kp^2\right)
    \int \limits_0^{\frac{\pi}{2}} d\theta \; \sin\theta\,\cos^2\!\theta \;
    I_0\left(2w^2\frac{\omega}{c}\kp\sin\theta \right)
    \exp\left[-w^2\frac{\w^2}{c^2}\sin^2\theta\right],\label{eq:17b}
\end{eqnarray}
\label{eq:17}
\end{subequations}
\end{widetext} 
and $I_0(z)$ is the modified Bessel function of the first kind and
zero order. In passing, it should be noted that in the large beam width
limit, for which the beam approaches a plane wave, it follows from
Eq.~(\ref{eq:17a}) that $p_{inc}=S\,\aokp$ where $S$ is the area of
the plane $x_3=0$ covered by the rough surface.

\subsection{Mean Differential Reflection Coefficient}

The differential reflection coefficient is defined as the fraction of
the total time-averaged flux incident on the surface that is scattered
into the element of solid angle $d\Omega_s$ about the scattering
direction $(\theta_s,\phi_s)$.  Since we are concerned with scattering
from a randomly rough surface, it is the averaged (or mean) of this
quantity over an ensemble of realizations of the surface that we need
to calculate. From its definition, we find from Eqs.~(\ref{Psc})
and (\ref{eq:16}) that the mean differential reflection coefficient
for the scattering of incident light of $\alpha$ polarization into
light of $\beta$ polarization is given by
\begin{eqnarray}
  \left\la \frac{\p R_{\beta\alpha}}{\p\Omega_s}\right\ra 
    &=& 
  \frac{1}{4\pi^2} \left(\frac{\w}{c}\right)^3 \cos^2\theta_s 
  \frac{ 
         \left< \left| {\mathcal E}_{\beta}(\bq_+ ,\w )\right|^2 \right>
       }{
          p_{inc}}.
        \label{eq:18}
         \qquad  
\end{eqnarray}

If we write the scattering amplitude ${\mathcal E}_{\beta}(\bq_+ ,\w
)$ as the sum of its mean value and the fluctuation about the mean,
\begin{widetext} 
\begin{eqnarray}
  {\mathcal E}_{\beta}(\bq_+ ,\w ) &=& 
  \la {\mathcal E}_{\beta}(\bq_+ ,\w )\ra
  + \left[{\mathcal E}_{\beta}(\bq_+ ,\w )- 
        \la {\mathcal E}_{\beta}(\bq_+ ,\w )\ra \right],
\end{eqnarray}
each term contributes separately to the mean differential reflection
coefficient 
\begin{subequations}
\begin{eqnarray}
  \left\la \frac{\p R_{\beta\alpha}}{\p\Omega_s}\right\ra 
  &=& \frac{1}{4\pi^2} \left(\frac{\w}{c}\right)^3 \cos^2\theta_s 
  \frac{\left\la  \left|{\mathcal E}_{\beta}(\bq_+ ,\w )\right|^2\right\ra}{p_{inc}} 
       \label{eq:34a} \\
   &=& \frac{1}{4\pi^2} \left(\frac{\w}{c}\right)^3 \cos^2\theta_s 
           \frac{ \left|\Big< {\mathcal E}_{\beta}(\bq_+ ,\w
                 )\Big> \right|^2}{p_{inc}}
       + \frac{1}{4\pi^2} \left(\frac{\w}{c}\right)^3 \cos^2\theta_s 
          \frac{   \Big<  \left| {\mathcal E}_{\beta}(\bq_+ ,\w ) \right|^2 \Big> 
                   - \left| \Big< {\mathcal E}_{\beta}(\bq_+,\w) \Big> \right|^2 
               }{
                   p_{inc}}. 
    \label{eq:34b}
  \end{eqnarray}
\label{eq:34}
\end{subequations}
\end{widetext}  
The first term in Eq.~(\ref{eq:34b}) gives the contribution to
the mean differential reflection coefficient from the light that has
been scattered coherently,
\begin{eqnarray}
  \left\la \frac{\p R_{\beta\alpha}}{\p\Omega_s}\right\ra _{coh}
      &=& \frac{1}{4\pi^2} \left(\frac{\w}{c}\right)^3 \cos^2\theta_s 
        \frac{ \left|\left \la {\mathcal E}_{\beta}(\bq_+ ,\w )\right\ra \right|^2}{p_{inc}}.
        \qquad    \label{eq:35}
\end{eqnarray}
The second term gives the contribution to the mean differential
reflection coefficient from the light that has been scattered
incoherently,
\begin{align}
 &\left\la \frac{\p R_{\beta\alpha}}{\p\Omega_s}\right\ra _{incoh}
  = \frac{1}{4\pi^2} \left(\frac{\w}{c}\right)^3 \cos^2\theta_s
  \nonumber \\
  & \quad \qquad \qquad 
  \times 
            \frac{   \left\la  \left| {\mathcal E}_{\beta}(\bq_+ ,\w ) \right|^2 \right\ra 
                   - \left| \Big< {\mathcal E}_{\beta}(\bq_+,\w) \Big> \right|^2 
               }{
                   p_{inc}}. 
                 \label{eq:36}
\end{align}

The dependencies of the right-hand sides of these expressions on the
polarization index $\alpha$ is through the dependence of the
amplitudes ${\mathcal E}_{\beta}(\bq_+,\w)$ on the surface current ${\bf
  J}_H(\bxp|\w)$ in Eqs.~(\ref{eq:8}). This surface current satisfies
the inhomogeneous integral equations, Eqs.~(\ref{eq:5}), in which the
inhomogeneous terms depend on the incident field, and hence on its
polarization $\alpha = p, s$.  Thus ${\mathcal E}_{\beta}(\bq_+ ,\w )$
depends implicitly on the polarization $\alpha$ of the incident field,
and so therefore does the differential reflection coefficient.

The procedure now is to generate a large number $N_p$ of realizations
of the surface profile function $\zxp$, and for each realization to
solve the scattering problem for an incident field of $p$ or $s$
polarization. The solution is then used to calculate the scattering
amplitude ${\mathcal E}_{\beta}(\bq_+,\w)$ and $|{\mathcal
  E}_{\beta}(\bq_+,\w)|^2$. An arithmetic average of the $N_p$ results
for these quantities yields the quantities $\left|\la {\mathcal
    E}_p(\bq_+,\w)\ra \right|^2$ and $\la |{\mathcal
  E}_s(\bq_+,\w)|^2\ra $ entering Eqs.~(\ref{eq:35})--(\ref{eq:36})
for the mean differential reflection coefficient.

\subsection{Energy conservation}

To facilitate the discussion of the conservation of energy, let us
define the following quantity
\begin{eqnarray}
  {\cal U}^\beta_\alpha( \theta_0,\phi_0) 
     &=& 
  \int \! d\Omega_s \; \left\la \frac{\p R_{\beta\alpha}}{\p\Omega_s}\right\ra.
  \label{eq:Energy-conservation-A}
\end{eqnarray}
Recalling the definition of the mean differential reflection
coefficient, it follows that the physical significance of ${\cal
  U}^\beta_\alpha( \theta_0,\phi_0)$ is that it is the fraction of the
flux of the incident $\alpha$-polarized light that is scattered into
$\beta$-polarized light by the rough surface irrespective of
scattering direction.

For a perfectly conducting surface, {\em all} power flux incident onto
the rough surface has to be converted into scattered power flux
leaving the surface, since there is no absorption in the
system. Hence, this is nothing but energy conservation, and it can be
expressed in terms of ${\cal U}^\beta_\alpha( \theta_0,\phi_0)$ as
\begin{eqnarray}
  {\cal U}(\theta_0,\phi_0) 
  &=&  \sum_{\alpha=p,s} {\cal U}_\alpha(\theta_0,\phi_0)
  \nonumber \\
  &=& \sum_{\beta=p,s} \sum_{\alpha=p,s} 
      {\cal U}^\beta_\alpha(\theta_0,\phi_0)     
  \nonumber \\
  &=& 1,
  \label{eq:Energy-conservation}
\end{eqnarray}
where the $\alpha$-summation over the polarization of the incident light
is only non-trivial in cases where the incident beam does not
have a well-defined $p$- or $s$-polarization. It was pointed out in
the previous subsection, that the mean differential reflection
coefficient can be separated into a coherent and an incoherent
component. The same applies therefore to ${\cal U}((\theta_0,\phi_0)$
and related quantities.

We note that Eq.~(\ref{eq:Energy-conservation}) is rather useful for
estimating the quality of the simulations, including making sure that the
discretization interval is fine enough. However, it should be stressed that
relation (\ref{eq:Energy-conservation}) is only a necessary condition,
and its satisfaction does not guarantee that the
simulations are correct.

\section{Results and Discussions}
\label{Sec:Results}

We have carried out calculations of the scattering of $p$- and
$s$-polarized light from a randomly rough perfectly conducting surface
with an rms height $\delta=\lambda$ and a transverse correlation
length $a=2\lambda$, where $\lambda$ is the wavelength of the incident
field. The polar angles of incidence are $\theta_0=0^{\circ}$,
$20^{\circ}$ and $40^{\circ}$, while the azimuthal angle of incidence
in all cases is $\phi_0=0^\circ$. The surface is generated at a
$112\times112$ grid of points covering an area $L^2=16\lambda\times
16\lambda$. The integration mesh size is therefore $\Delta
x=\lambda/7$. The calculations were carried out for an incident field
in the form of a Gaussian beam
[Eqs.~(\ref{eq:11})] of width $w=4\lambda$.

\begin{table}[tbh]
\begin{center}
\begin{tabular}{cccccccccc}
\hline
\hline
$\theta_0$ [deg]                               \qquad  & 
$\alpha$                                       \qquad & \quad &
${\cal U}$                         &
${\mathcal U}_\text{incoh}$         \qquad &  
${\cal U}_\text{coh} [10^{-4}]$     \qquad  &
${\cal U}^p_\alpha/{\cal U}$        \qquad  & \quad  &   
${\cal U}^s_\alpha/{\cal U}$        \qquad    \\ 
\hline
  $0$  & $p$ && 0.9976 & 0.9975 & 0.9 & 0.5054  && 0.4946 \\
  $20$ & $p$ && 0.9962 & 0.9961 & 0.9 & 0.5315  && 0.4686 \\
  $40$ & $p$ && 0.9951 & 0.9947 & 3.8 & 0.5407  && 0.4592 \\*[1.5mm] 
  $0$  & $s$ && 0.9970 & 0.9967 & 3.1 & 0.5021  && 0.4979 \\
  $20$ & $s$ && 0.9966 & 0.9963 & 2.8 & 0.4939  && 0.5061 \\ 
  $40$ & $s$ && 0.9953 & 0.9948 & 4.9 & 0.4834  && 0.5166 \\
\hline
\hline
\end{tabular}
\end{center}

\caption{\label{tab:2} 
  The energy conservation for various polar angles of
  incidence ($\theta_0$) and incidence polarizations ($\alpha$) 
  for the surface parameters
  given in the text. The surface and scattering amplitude were
  discretized on $112 \times 112$ and $101\times 101$ grids,
  respectively.  These results were obtained on the basis of
  Eqs.~(\protect\ref{eq:Energy-conservation-A}) and
  (\ref{eq:Energy-conservation}).}
\end{table}

\begin{figure*}[tbh]
  \begin{center}
    \includegraphics*[width=0.75\columnwidth]{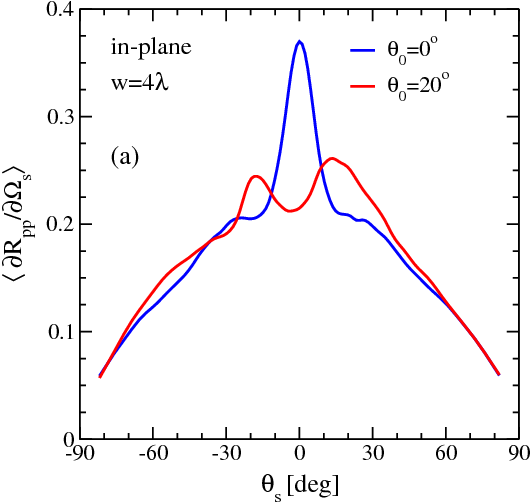} 
    \qquad \quad
    \includegraphics*[width=0.75\columnwidth]{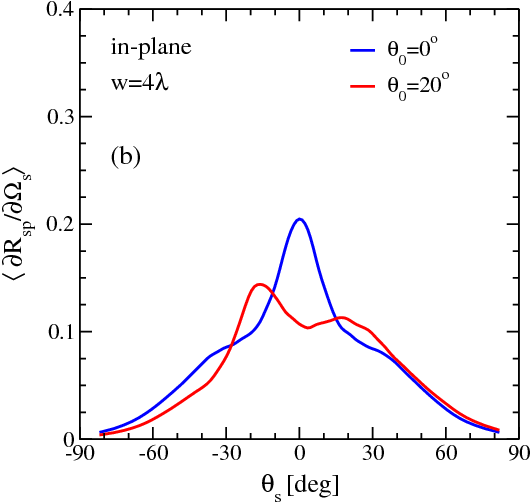}
    \\*[3mm]
    \includegraphics*[width=0.75\columnwidth]{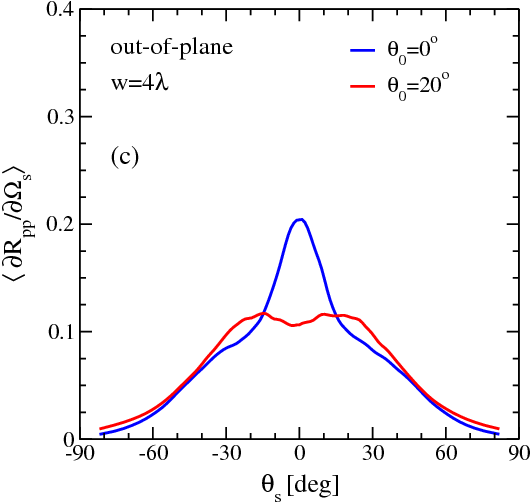}
    \qquad \quad
    \includegraphics*[width=0.75\columnwidth]{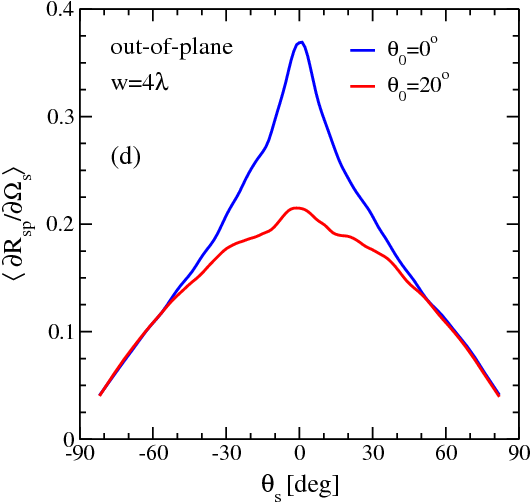}
  \end{center}
  \caption{(Color online) The mean differential reflection
    coefficients, $\left<\partial
      R_{\beta\alpha}/\partial\Omega_s\right>$ $(\alpha \rightarrow
    \beta$), as functions of the polar scattering angle $\theta_s$ for
    the in-plane ($\phi_s=\phi_0$ or $\phi_s=\phi_0+180^{\circ}$) and
    out-of-plane ($\phi_s= \phi_0 \pm 90^{\circ}$), co- ($p\rightarrow
    p$) and cross-($p\rightarrow s$) polarized scattering of a
    $p$-polarized incident beam ($\alpha=p$) of width $w=4\lambda$
    ($\theta_0=0^{\circ}$ and
    $\theta_0=20^{\circ}$; 
    $\phi_0=0^{\circ}$) scattered from a Gaussian randomly rough
    perfectly conducting surface.  The Gaussian correlated surface had
    a correlation length $a=2\lambda$ and an rms height
    $\delta=\lambda$. To facilitate comparison between the various
    configurations presented in this figure, notice that we have used
    similar scales for all ordinate axes. Moreover, to simplify the
    presentation of the figures, a convention was adopted where
    negative~(positive) values of $\theta_s$ correspond to
    $\phi_s=\phi_0+180^\circ$ ($\phi_s=\phi_0$).}
  \label{Fig1}
\end{figure*}

In Fig.~\ref{Fig1} we plot the mean differential reflection
coefficients as functions of the polar scattering angle $\theta_s$ for
the in-plane ($\phi_s=0^{\circ}$) and out-of-plane
($\phi_s=\pm 90^{\circ}$), co- ($p\rightarrow p$) and cross-($p\rightarrow
s$) polarized scattered light due to a $p$-polarized Gaussian beam
incident on the surface. The results depicted in Figs.~\ref{Fig1} were
obtained as averages over $12,000$ realizations of the surface profile
function.  In obtaining these results we have noted that at least for
the roughness parameters we have assumed, the contribution to the mean
differential reflection coefficient from the light scattered
coherently is smaller than the contribution from the light scattered
incoherently by a factor of approximately $10^{-4}$ (see
Table~\ref{tab:2} for details).

\begin{figure*}[th]
  \begin{center}
    \includegraphics*[width=0.75\columnwidth]{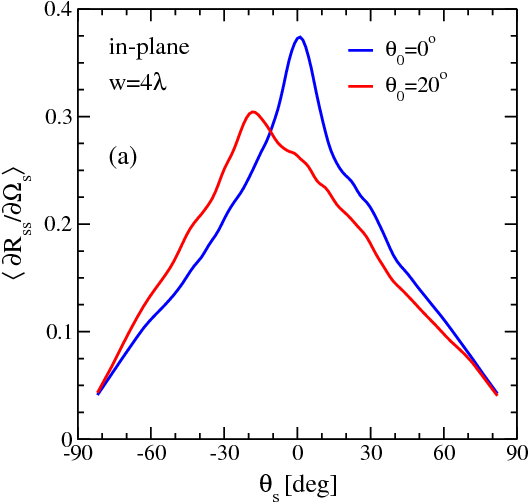} 
    \qquad \quad
    \includegraphics*[width=0.75\columnwidth]{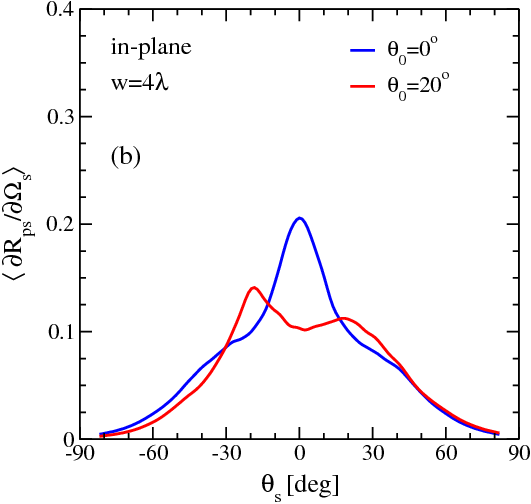}
    \\*[3mm]
    \includegraphics*[width=0.75\columnwidth]{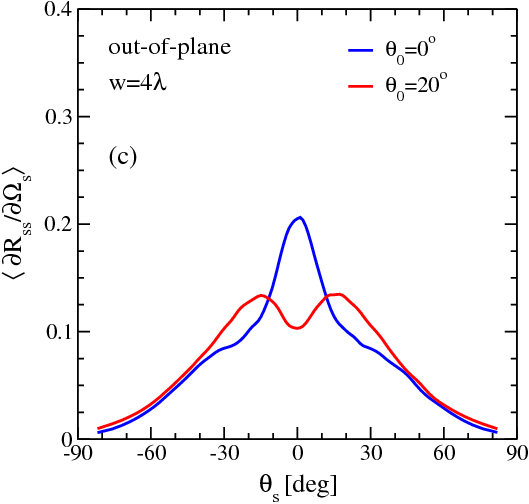}
    \qquad \quad
    \includegraphics*[width=0.75\columnwidth]{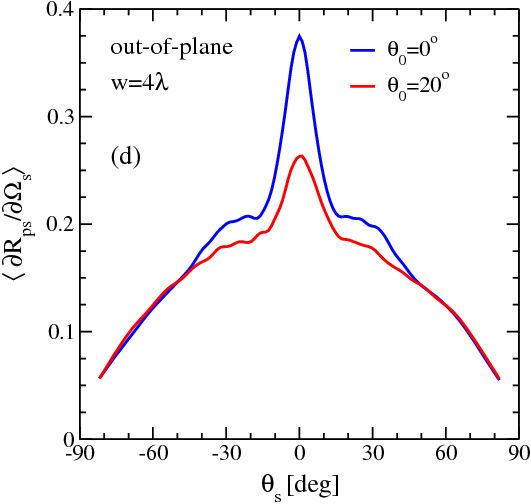}
  \end{center}
  \caption{Same as Fig.~\protect\ref{Fig1}, but for an 
    $s$-polarized incident beam.}
  \label{Fig2}
\end{figure*}

There is no single scattering contribution in the cases of in-plane
cross-polarized [Fig.~\ref{Fig1}(b)] and out-of-plane co-polarized
[Fig.~\ref{Fig1}(c)] scattering. This we believe is the main reason
for the reduced amplitude of the mean differential reflection
coefficients in these cases relative to those of Fig.~\ref{Fig1}(a)
and (d) where single scattering is allowed.  The peaks at
$\theta_s=0^{\circ}$ and $-20^{\circ}$ 
\footnote{When in the text discussing the results of Figs.~\ref{Fig1}
  and \ref{Fig2}, we follow the sign convention for $\theta_s$
  introduced in the caption of Fig.~\ref{Fig1}. Elsewhere, however,
  the standard spherical coordinate convention ($\theta_s\geq 0^\circ$)
  will be followed.}  for in-plane co-polarized scattering
[Figs.~\ref{Fig1}(a)] are enhanced backscattering
peaks~\cite{EBP-1,EBP-2,EBP-3,Elena}.  However, the structures seen as
peaks in the backscattering directions of the cross-polarized
scattering, Fig.~\ref{Fig1}(b), are not real peaks, as will be seen
below from the full angular intensity distributions. The results that
the mean differential reflection coefficients for out-of-plane co- and
cross-polarized scattering [Figs.~\ref{Fig1}(c) and (d)] are even
functions of $\theta_s$ are consequences of the scattering geometry,
namely that $\phi_0=0^{\circ}$, $\phi_s=\pm90^{\circ}$, and the
isotropy of the power spectrum of the surface roughness.

In Fig.~\ref{Fig2} we present corresponding results to those of
Fig.~\ref{Fig2}, but now for an $s$-polarized incident Gaussian beam.
There is no single scattering contribution to the in-plane
cross-polarized and out-of-plane co-polarized scattering, as in the
case of $p$ polarization.  Also in this case the peaks seen in the
in-plane co-polarized scattering [Fig.~\ref{Fig2}(a)] are enhanced
backscattering peaks, while the structures seen in the in-plane
cross-polarized scattering [Fig.~\ref{Fig2}(b)] in the backscattering
direction are not real peaks.

\begin{figure*}[tbh]
  \begin{center}
    \includegraphics[width=0.9\columnwidth]{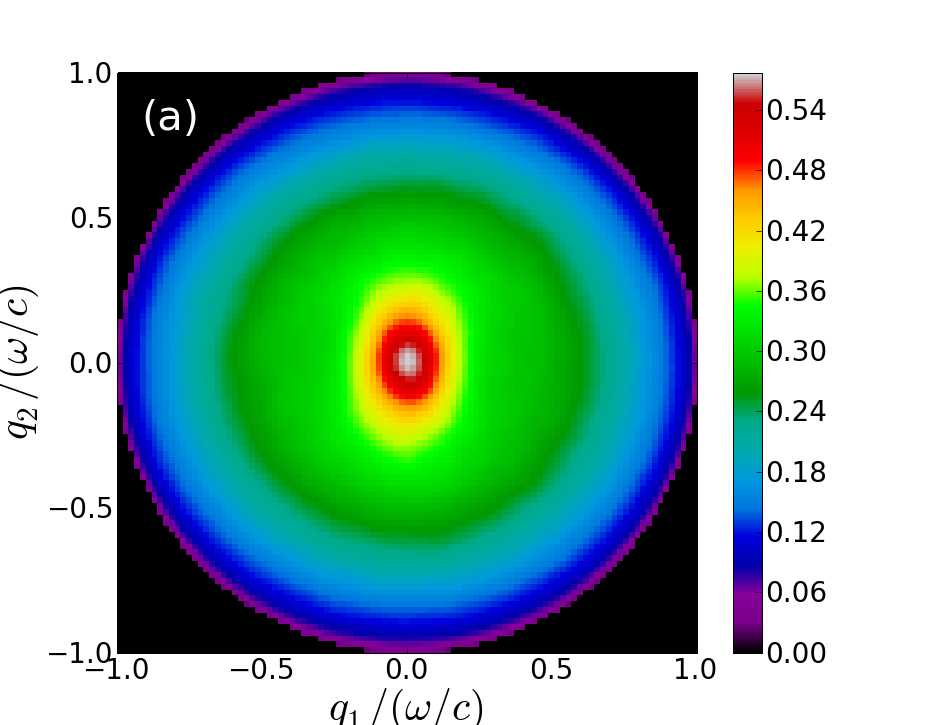} \qquad
    \includegraphics[width=0.9\columnwidth]{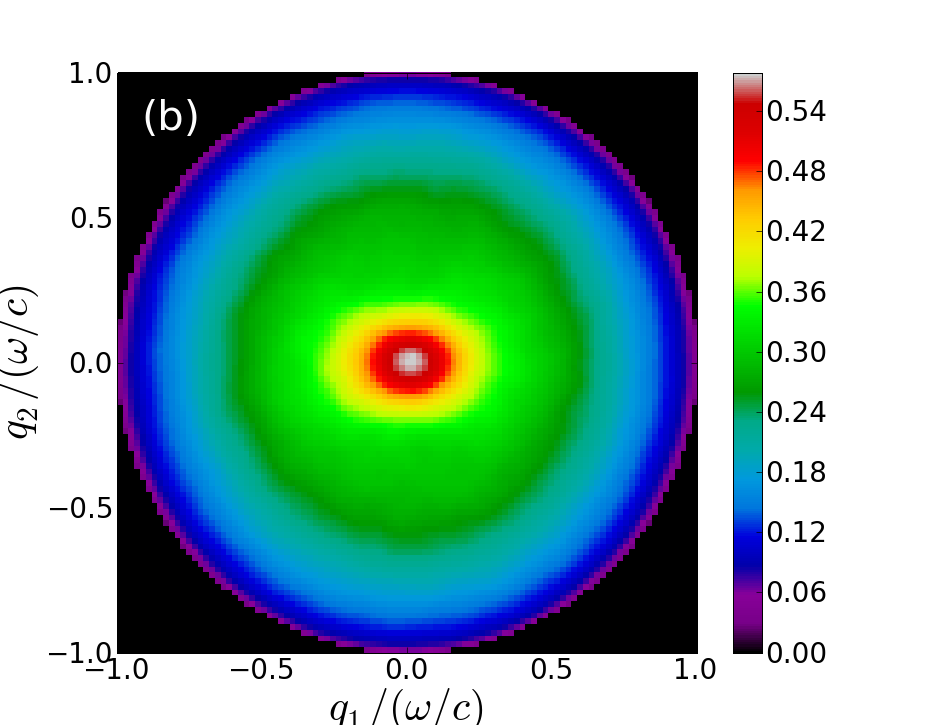} \\
    \includegraphics[width=0.9\columnwidth]{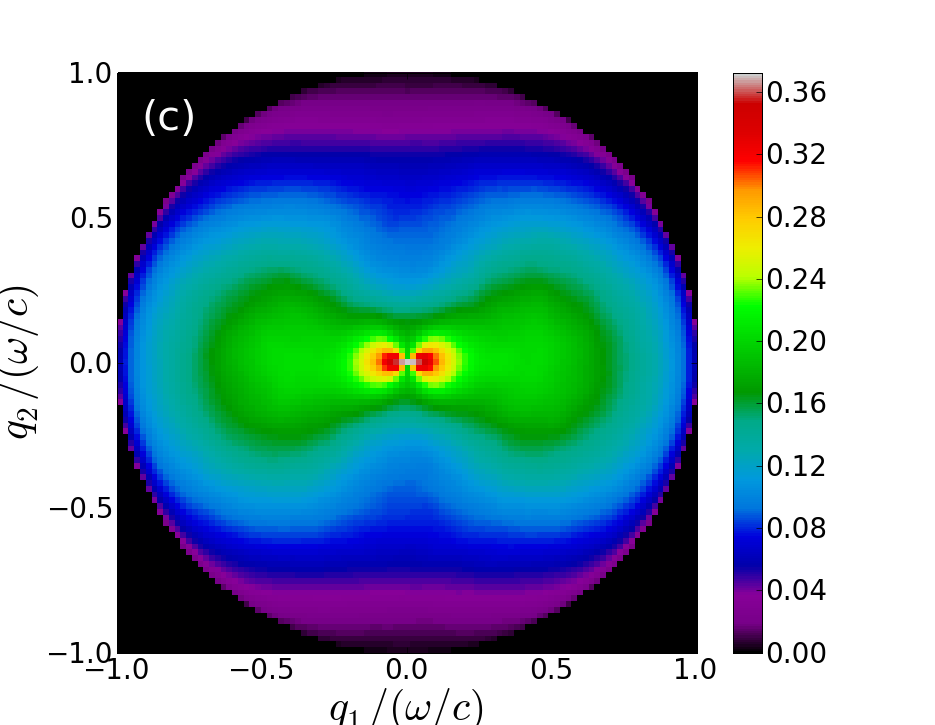}  \qquad
    \includegraphics[width=0.9\columnwidth]{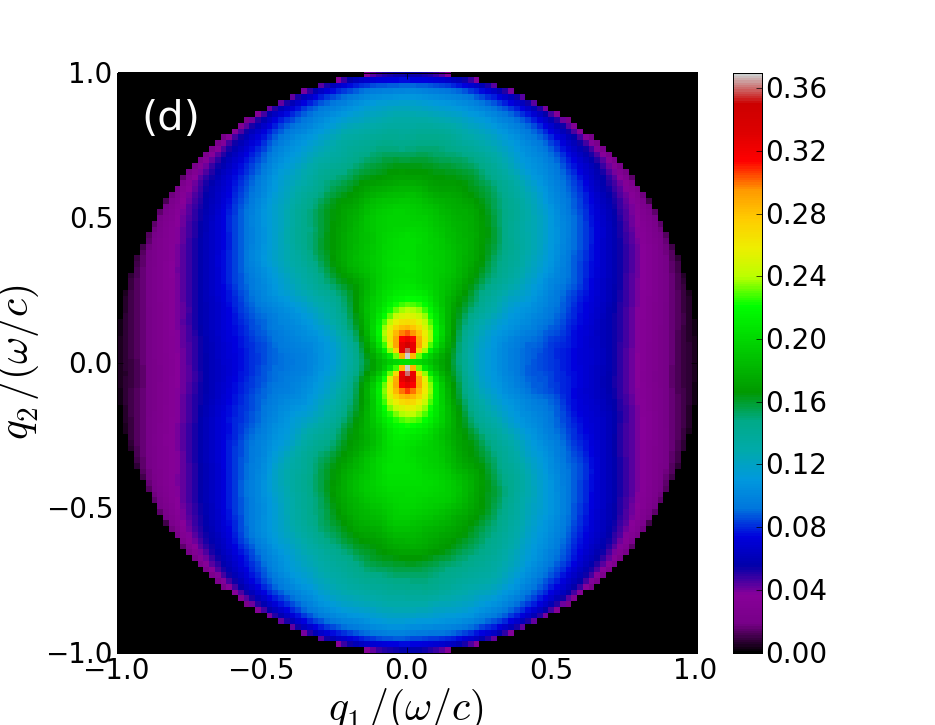}  \\
    \includegraphics[width=0.9\columnwidth]{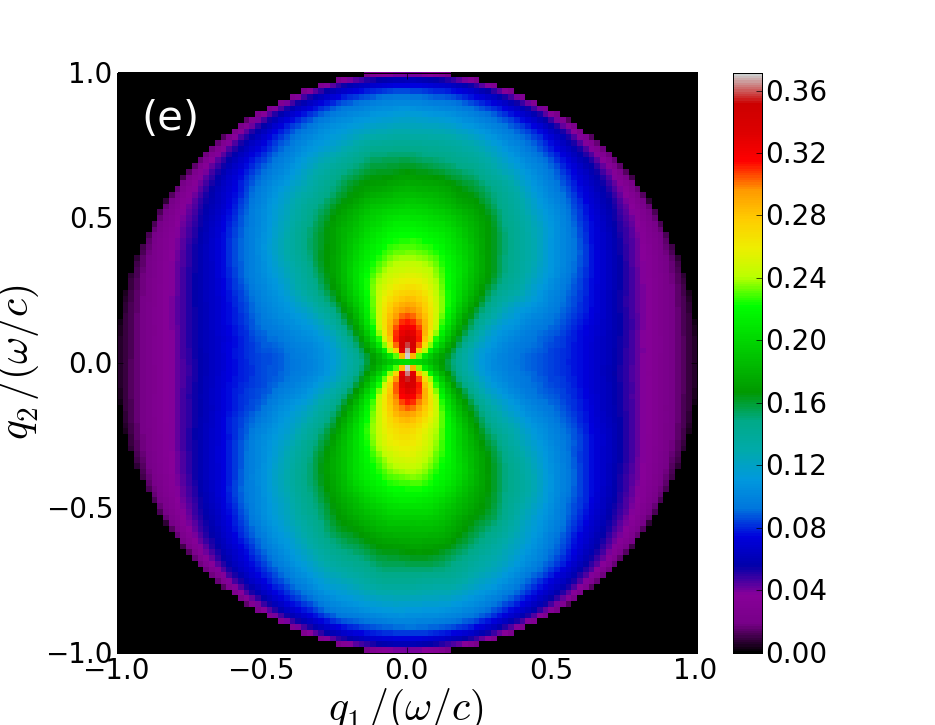}  \qquad
    \includegraphics[width=0.9\columnwidth]{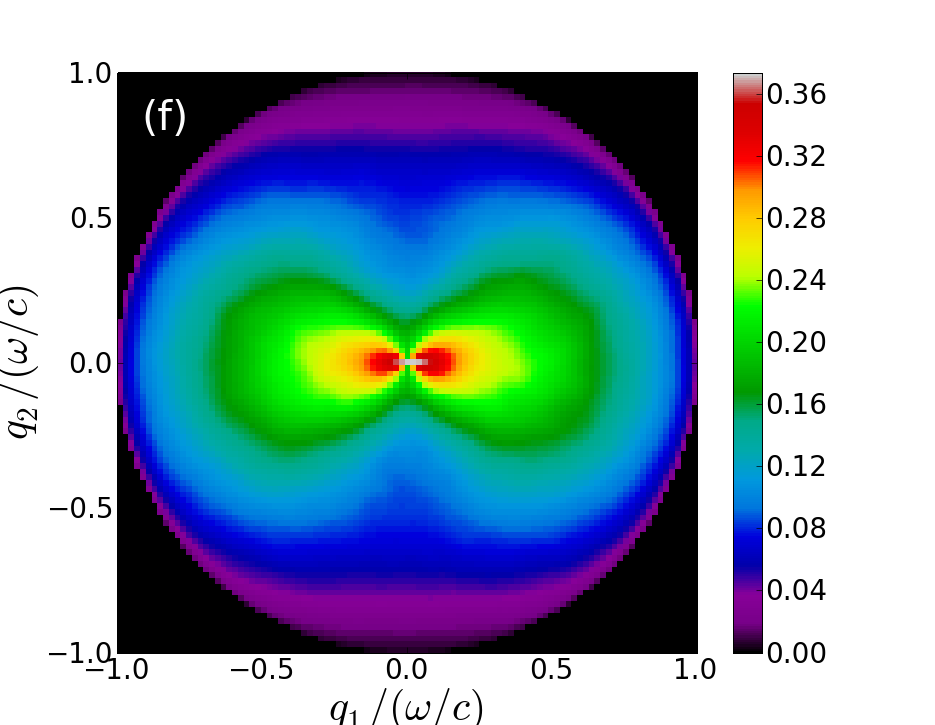}  
  \end{center}
  \caption{(Color online) The complete angular distributions of the
    mean differential reflection coefficient, $\left<\partial
      R_{\beta\alpha}/\partial\Omega_s\right>$, for the scattering of
    an $\alpha$-polarized Gaussian beam incident on the surface at
    polar angle $\theta_0=0^\circ$ and azimuthal angle
    $\phi_0=0^\circ$.  The perfectly conducing rough surface was
    characterized by a Gaussian height distribution of rms-value
    $\delta=\lambda$ and a Gaussian correlation function of transverse
    correlation length $a=2\lambda$.  The incident beam was
    $p$-polarized in Figs.~\protect\ref{Fig_Inc_00}(a), (c) and (e)
    [left column]; and $s$-polarized in
    Figs.~\protect\ref{Fig_Inc_00}(b), (d) and (f) [right
    column]. Moreover, in the top two figures
    [Figs.~\protect\ref{Fig_Inc_00}(a) and (b)] the polarization of
    the scattered light was not recorded; in
    Figs.~\protect\ref{Fig_Inc_00}(c) and (d) [central row] only
    $p$-polarized scattered light was recorded; while the bottom two
    figures correspond to recording only $s$-polarized scattered light
    [Figs.~\protect\ref{Fig_Inc_00}(e) and (f)].  The rough surface,
    covering an area $16\lambda \times 16\lambda$, was discretized at
    a grid of $112\times112$ points corresponding to a distcretization
    interval $\lambda/7$ for both directions. The presented figures
    were obtained by averaging the mean differential reflection
    coefficient over $12,000$ surface realizations.}
  \label{Fig_Inc_00}
\end{figure*}

\begin{figure*}[tbh]
  \begin{center}
    \includegraphics[width=0.9\columnwidth]{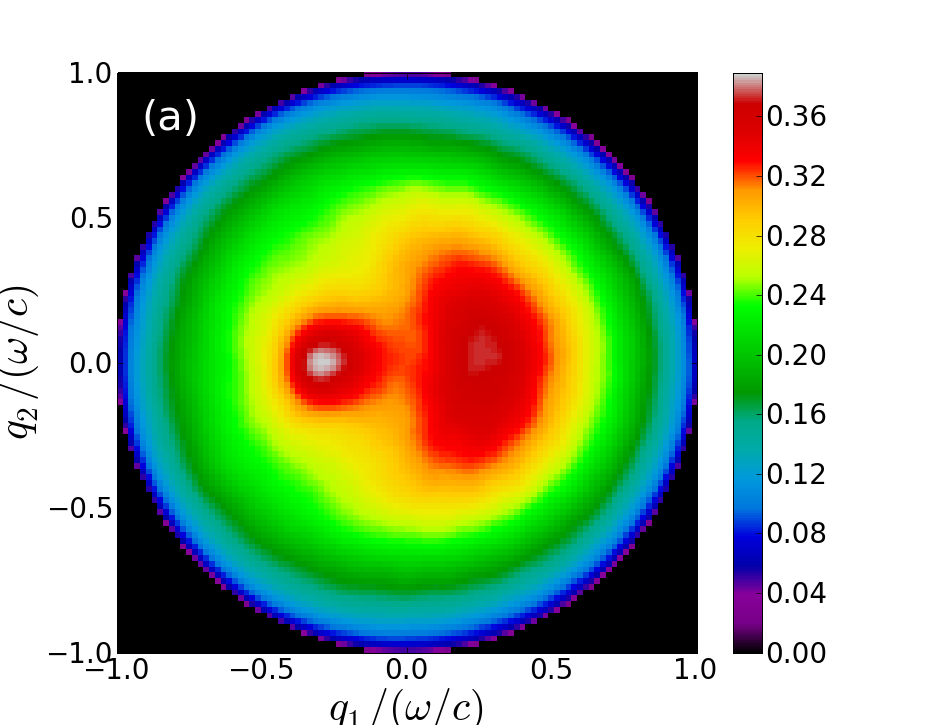} \qquad
    \includegraphics[width=0.9\columnwidth]{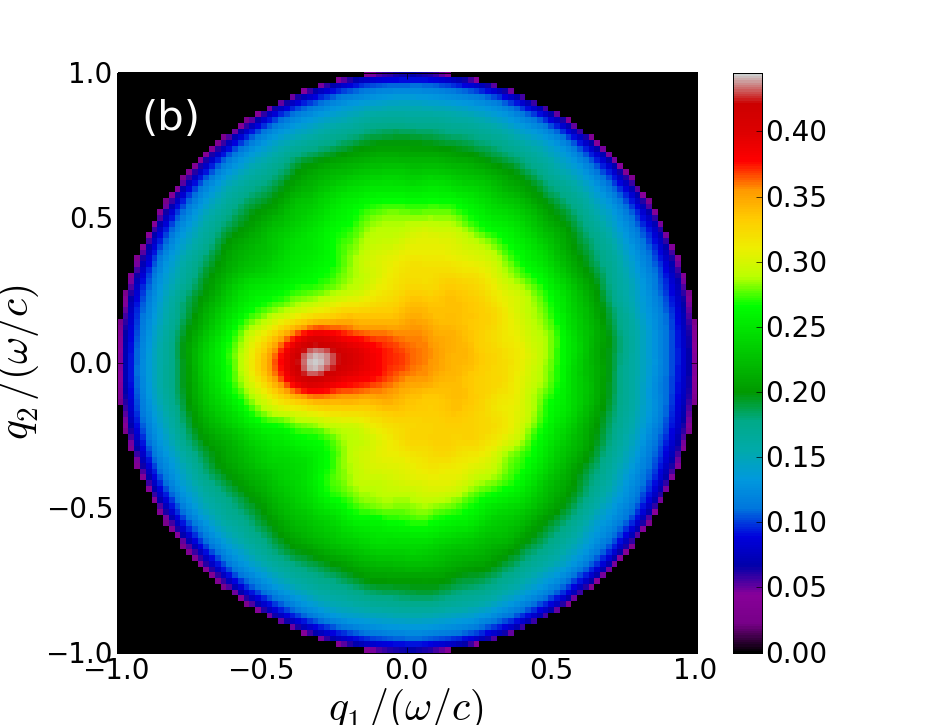} \\ 
    \includegraphics[width=0.9\columnwidth]{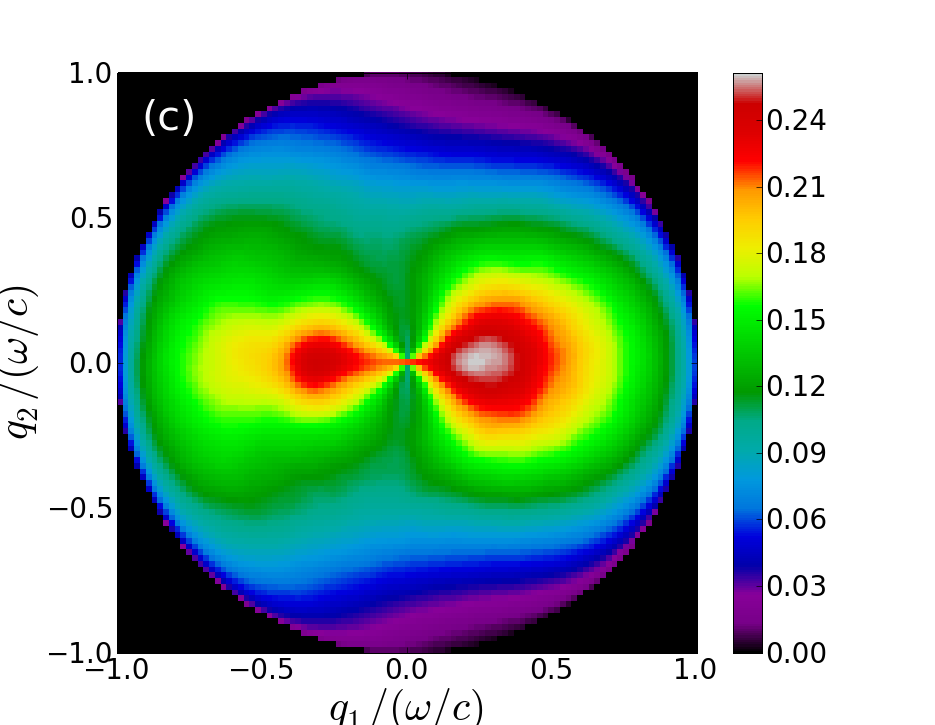}  \qquad
    \includegraphics[width=0.9\columnwidth]{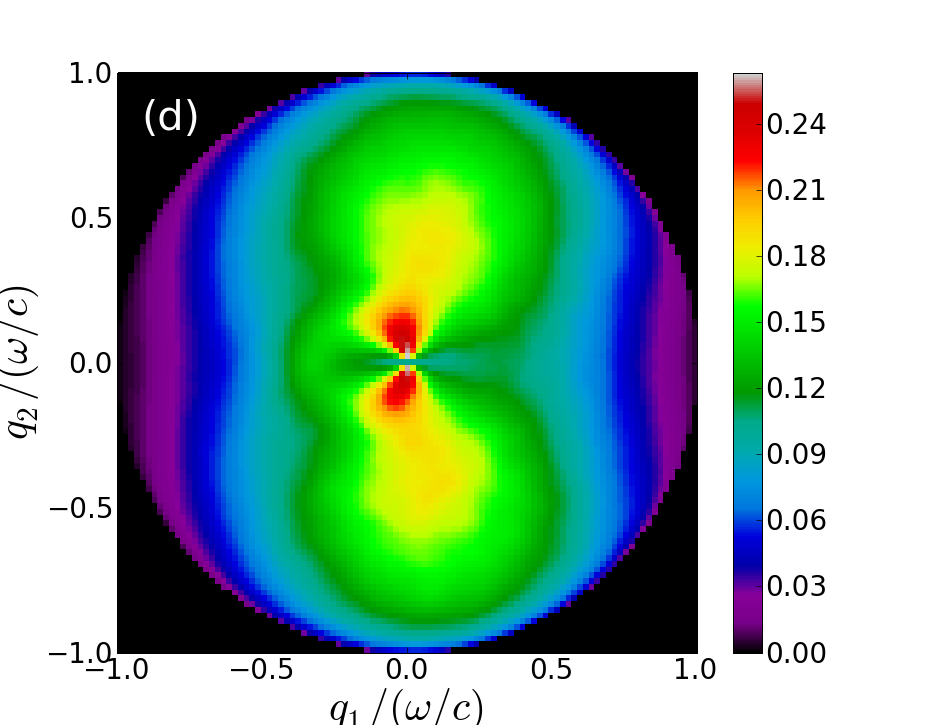}  \\
    \includegraphics[width=0.9\columnwidth]{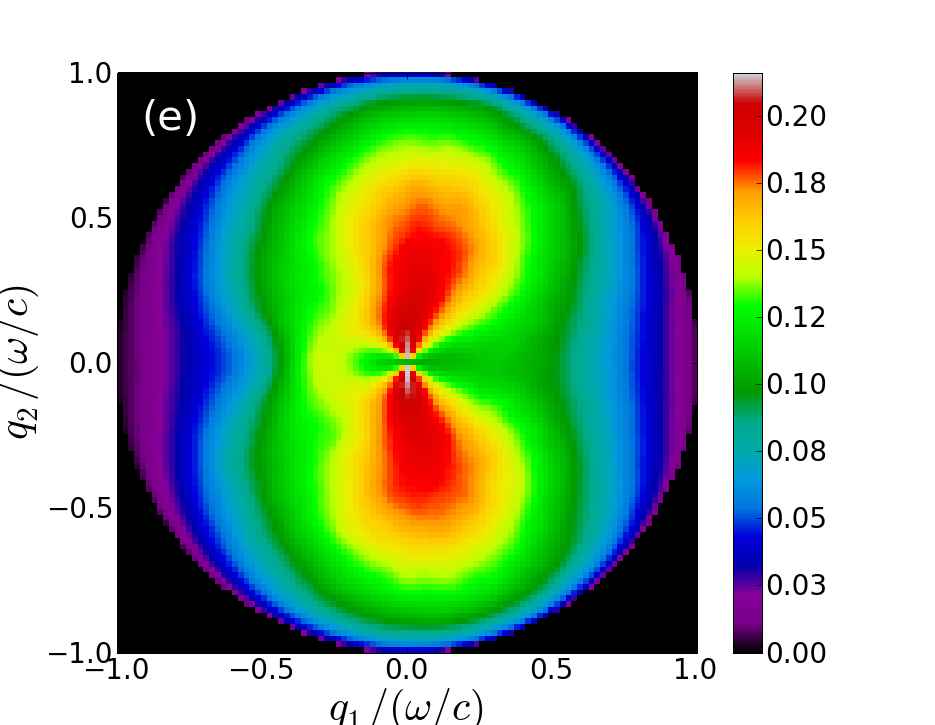}  \qquad
    \includegraphics[width=0.9\columnwidth]{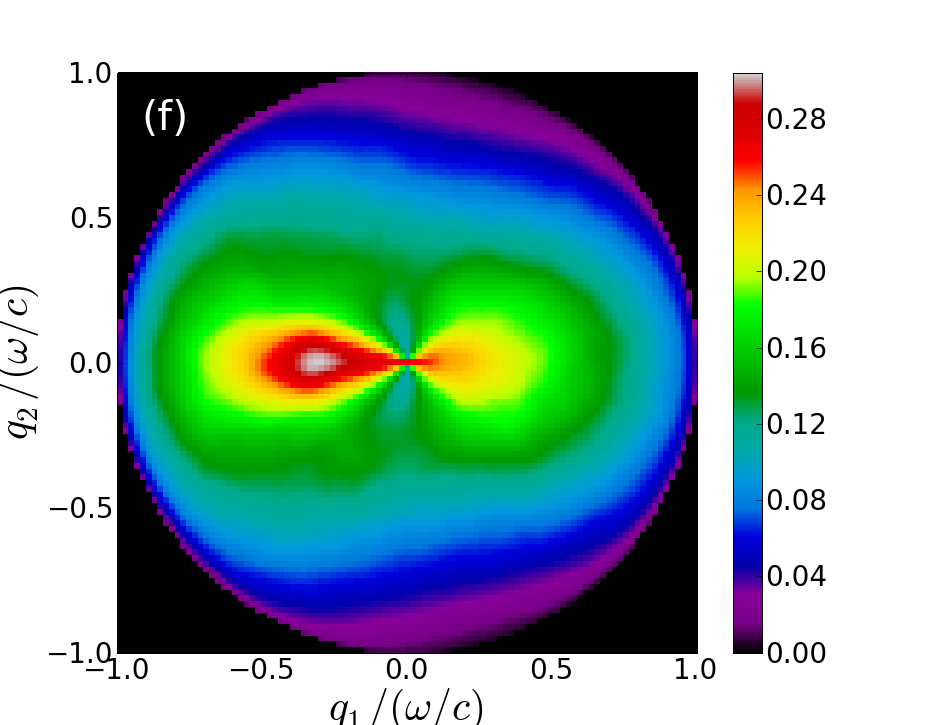}  
  \end{center}
  \caption{(Color online) Same as Figs.~\protect\ref{Fig_Inc_00}, but for a polar
    angle of incidence $\theta_0=20^\circ$.}
  \label{Fig_Inc_20}
\end{figure*}
\begin{figure*}[tbh]
  \begin{center}
    \includegraphics[width=0.9\columnwidth]{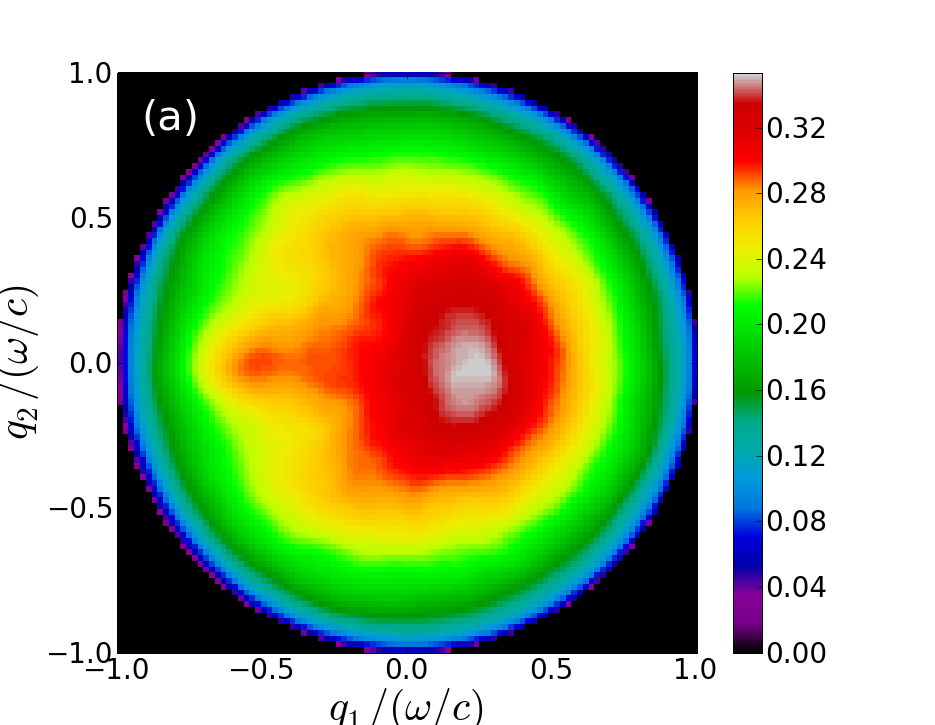} \qquad
    \includegraphics[width=0.9\columnwidth]{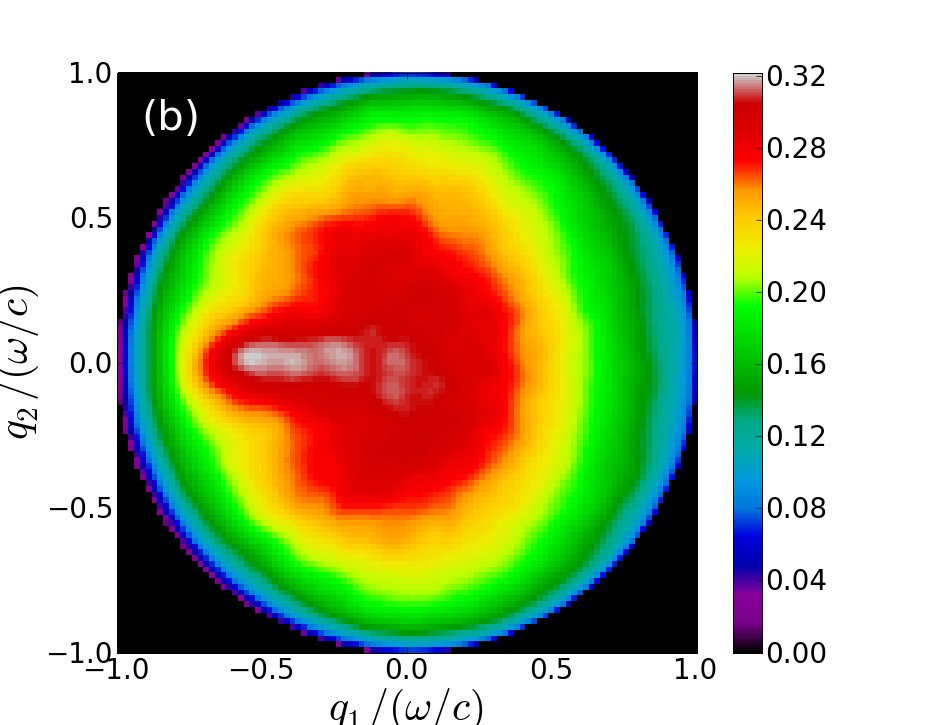} \\ 
    \includegraphics[width=0.9\columnwidth]{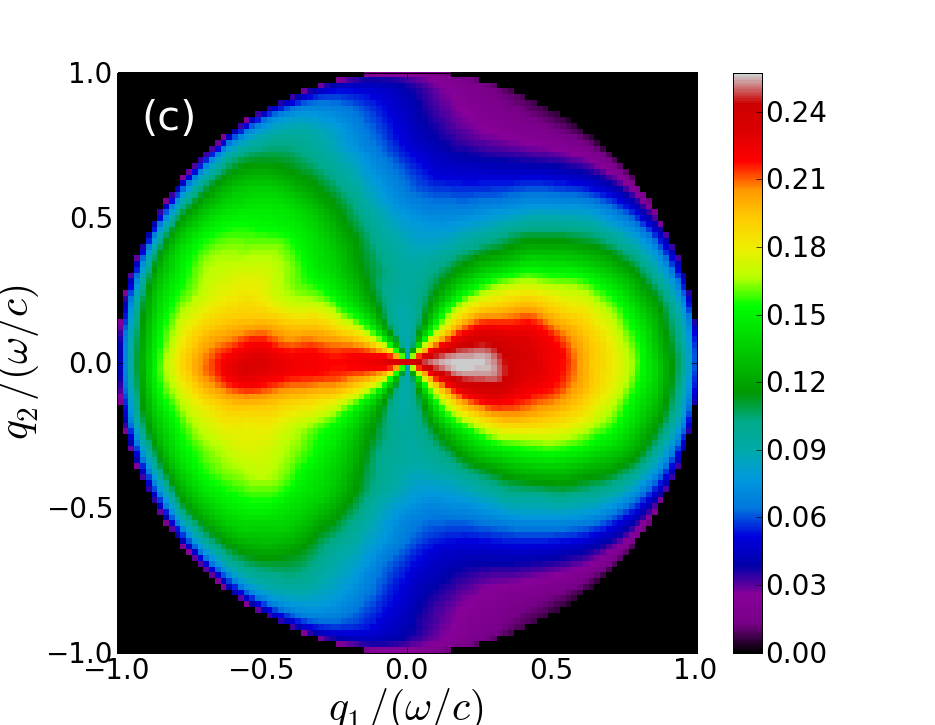}  \qquad
    \includegraphics[width=0.9\columnwidth]{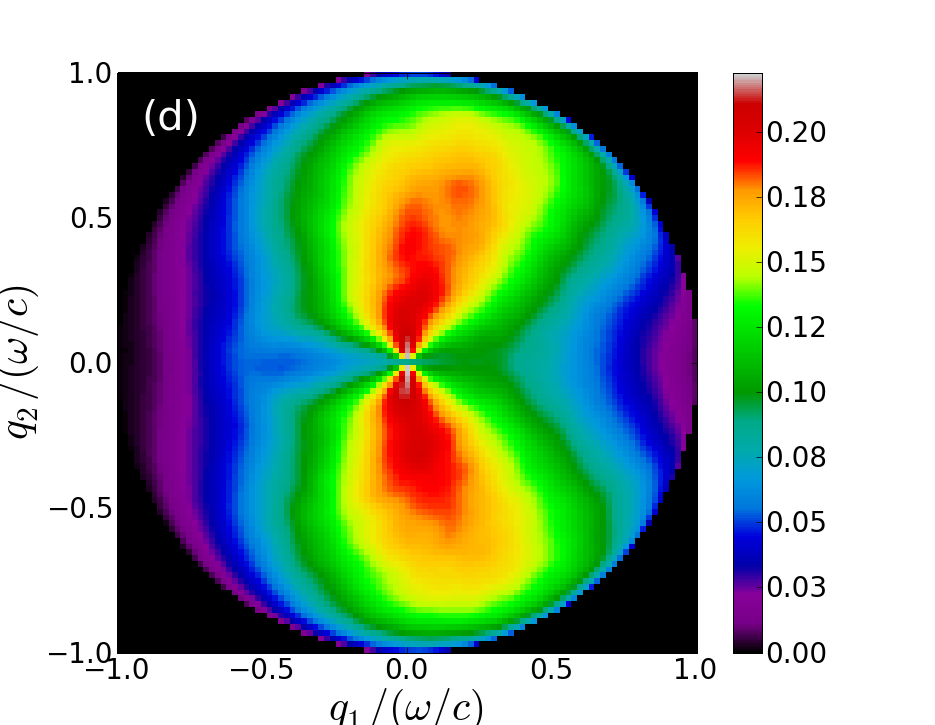}  \\
    \includegraphics[width=0.9\columnwidth]{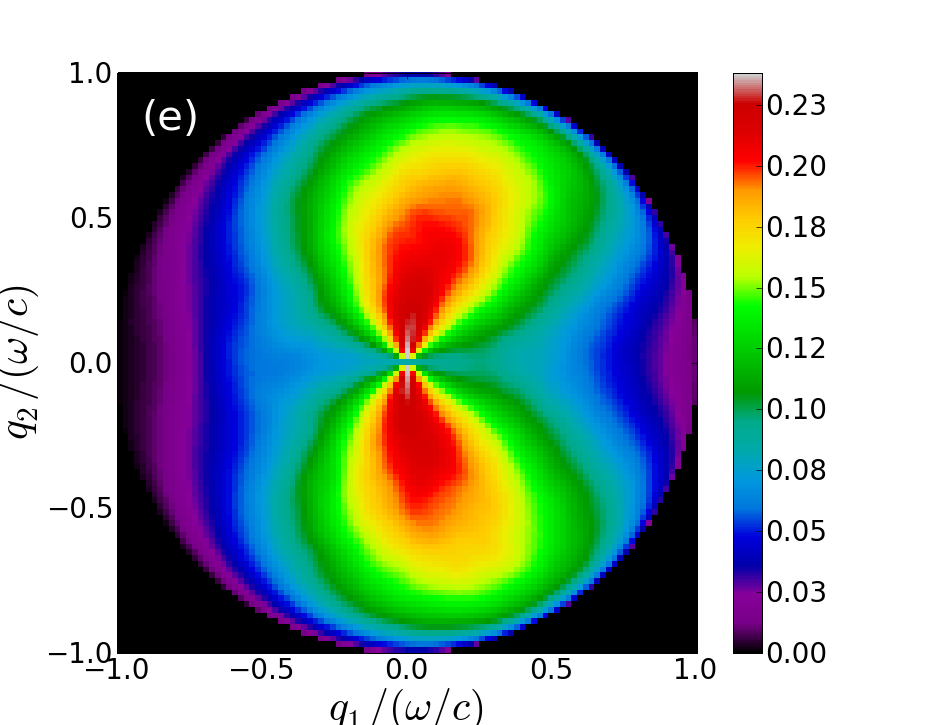}  \qquad
    \includegraphics[width=0.9\columnwidth]{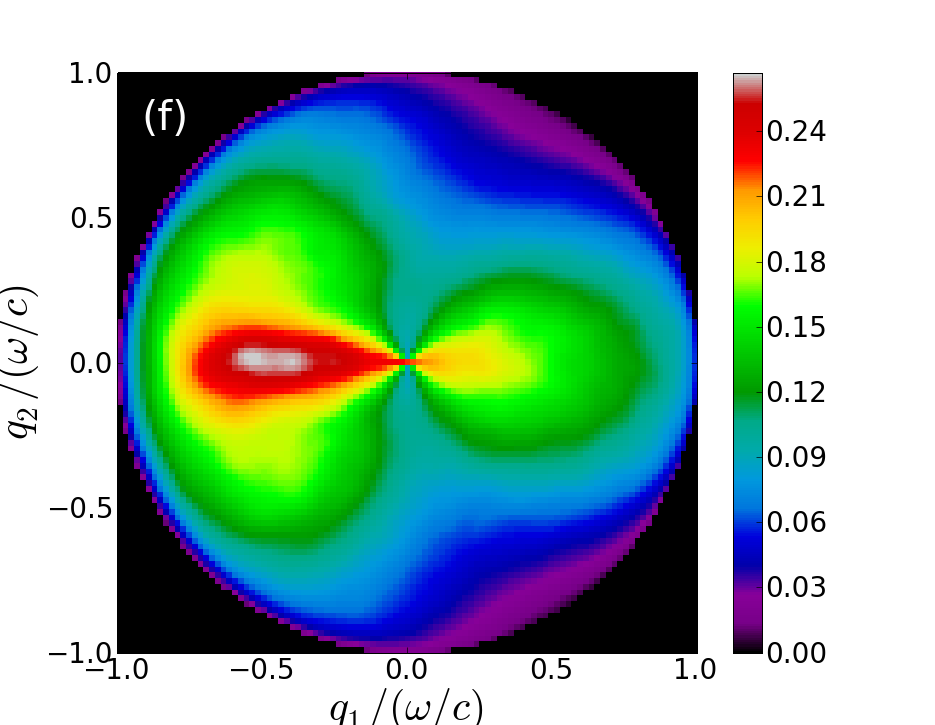}  
  \end{center}
  \caption{(Color online) Same as Figs.~\protect\ref{Fig_Inc_00}, but for a polar
    angle of incidence $\theta_0=40^\circ$.}
  \label{Fig_Inc_40}
\end{figure*}

The full angular distribution of the intensity of the scattered light
is presented as color contour plots in
Figs.~\ref{Fig_Inc_00}--\ref{Fig_Inc_40}, which correspond to the
polar angles of incidence $\theta_0=0^\circ$, $20^\circ$, and
$40^\circ$, respectively, and for several combinations of the
polarizations of the incident and scattered light~\footnote{Note that
  we from now and onwards will adapt standard spherical coordinates so
  that $\theta_s\geq 0^\circ$.}.
To the best of our
knowledge, this is the first time that the full angular distributions
of the light scattered from a strongly rough surface have been
obtained by a rigorous computer simulation approach.  It is observed
from Figs.~\ref{Fig_Inc_00}--\ref{Fig_Inc_40} that the angular
distributions, for given polarizations of the incident and scattered
light, are far from trivial, and show strong and complex angular
dependencies. With the full angular dependence of the scattered light
available, the energy conservation of the simulations performed can be
obtained by comparing the power incident on the surface to that being
scattered from it [see Eq.~(\ref{eq:Energy-conservation})]. For normal
incidence, we obtained ${\cal U}_p=0.9976$ and ${\cal U}_s=0.9970$ for
$p$- and $s$-polarized incident light, respectively. For the other
angles of incidence considered, $\theta_0=20^\circ$ and $40^\circ$,
energy conservation was satisfied within $0.5\%$ or better (see
Table~\ref{tab:2} for details).  Even if energy conservation is only a
necessary requirement, such results, however, still testify to the
accuracy of the simulations and the approaches used to obtain them.

It is interesting to note that for the roughness parameters
considered, the power in a normally incident beam is divided
essentially equally between $p$ and $s$ polarized scattered light
(independent of the polarization of the incident light). This effect
we attribute to multiple scattering.  For the other angles of
incidence, it is observed from Table~\ref{tab:2} that the fraction of
incident power being scattered into the same polarization as that of
the incident beam (co-polarized scattering), but still independent of
scattering direction, increases with the polar angle of incidence.

We will now discuss Figs.~\ref{Fig_Inc_00}--\ref{Fig_Inc_40} in more
detail: We start by considering the case of normal incidence;
$\theta_0=0^\circ$ and $\phi_0=0^{\circ}$~[Figs.~\ref{Fig_Inc_00}].
Recall that with the assumptions and conventions used in this work,
the electric field of an incident $p$-polarized Gaussian beam is in
the plane of incidence.  In Fig.~\ref{Fig_Inc_00}(a) we present a
contour plot of the mean differential reflection coefficient for the
scattering of $p$-polarized light into either $p$- or $s$-polarized
scattered light, ({\it i.e.} the polarization state of the scatted
light is not being recorded).  The angle-dependent scattering, in this
case, is for the most part rather isotropic, except for a slight
anisotropy seen as an elongated (along the $q_2$-direction) structure
around the normal scattering direction. This structure is caused by
the wider intensity distribution in the direction perpendicular to the
incident electric field as compared to the intensity distribution
along it.  The central peak present in Fig.~\ref{Fig_Inc_00}(a) is the
enhanced backscattering peak, and is not related to specular
scattering which for these roughness parameters can be neglected (see
Table~\ref{tab:2} for details). A similar behavior is seen for the
scattering of (normally) incident $s$-polarized light into either $p$-
or $s$-polarized light [Fig.~\ref{Fig_Inc_00}(b)]. Here an apparent
enhanced backscattering peak is also observed. In the case of
$s$-polarization, one sees though that the central anisotropic portion
of the scattering has a different orientation compared to that in the
case of $p$-polarization. It remains true, however, that there is a
stronger scattering perpendicular to the (average) direction of the
incident electric field independent of the polarization of the
incident light.

Based on these findings, one may be misled into believing that the
scattering for normal incidence into the two possible (linear)
polarizations, $p$ or $s$, is also more-or-less isotropic, except
maybe for some minor polarization dependence for the smaller
scattering angles $\theta_s$. However, this is rather far from being
true. In Figs.~\ref{Fig_Inc_00}(c) and (d) we present the scattering
into $p$-polarized scattered light from, respectively, a (normally)
incident $p$ and $s$ polarized Gaussian beam. Similarly, depicted in
Figs.~\ref{Fig_Inc_00}(e) and (f) are the scattering into
$s$-polarized scattered waves for an incident $p$- or $s$-polarized
Gaussian beam. We note that taking the sum of the distributions shown
in {\em e.g.}  Figs.~\ref{Fig_Inc_00}(c) and (e) produces the angular
distribution shown in Fig.~\ref{Fig_Inc_00}(a). From
Figs.~\ref{Fig_Inc_00}(c)--(f) it follows that the intensity
distributions for scattering from one polarization into another, or
into the same one, show a dipole-like angular dependence.

For co-polarized scattering, {\it i.e.} the polarization of the
incident light and the (recorded) polarization of the scattered light
are the same, the ``forward direction'' of the dipole-like pattern is
oriented along $q_1$ [Figs.~\ref{Fig_Inc_00}(c) and (f)], while for
cross-polarization, it is oriented along the $q_2$
direction~\footnote{The simulation results reported herein assumed an
  azimuthal angle of $\phi_0=0^\circ$ which also determines the
  directions of the electric field vector associated with the incident
  Gaussian beam, and also defines (in our convention) the rotation
  angle of the incident plane. Another choice for $\phi_0$ would
  consequently also alter the orientation of the dipole-like
  patterns.}.  For normal incidence, the ${\bf k}$-vector used to
define the incident Gaussian beam, does not (together with $\hat{\bf
  x}_3$) define a plane of incidence.  However, we have used the
convention in the simulations, that the plane of incidence is defined
as the plane having $\hat{\mbold{\phi}}_0=-\sin\phi_0 \hat{\bf q}_1
+\cos\phi_0 \hat{\bf q}_2$ as its normal vector which is well-defined
for all polar angles of incidence (also $\theta_0=0^\circ$) and
coincides with the usual definition when $\theta_0\neq 0$. Since
$\phi_0=0^\circ$ was assumed for all the simulation results presented,
it follows (with this convention) that the plane of incidence is the
$q_1q_3$-plane. With this definition for the plane of incidence, we
may rephrase the above observation: For co- and cross-polarized
scattering the dipole-like pattern is oriented {\em along} and {\em
  perpendicular} to the plane of incidence, respectively. Later we
will see that this statement also holds true for non-normal incidence.

It is noted that we have checked and found that the scattering of a
normally incident unpolarized beam by the rough surface, produces,
when both its $p$- and $s$-polarized components are recorded, a fully
rotationally symmetric intensity distribution (equal to the sum of the
distributions in Figs.~\ref{Fig_Inc_00}(a) and (b)).  If only $p$- or
$s$-polarized scattered light is recorded, one will still, with the
same type of unpolarized illumination, obtain rotationally symmetric
intensity distributions (equal to the sum of the distributions from
Figs.~\ref{Fig_Inc_00}(c) and (d), in the case of $p$-polarization, and
the sum of Figs.~\ref{Fig_Inc_00}(e) and (f) for $s$-polarization).

We now turn our attention to the scattering for non-normal incidence.
In Figs.~\ref{Fig_Inc_20} we present the results for the angular
distribution of the mean differential reflection coefficient for
either a $p$- or $s$-polarized Gaussian beam incident onto the surface
at a polar angle $\theta_0=20^\circ$ and scattered into various
polarization states.

From Figs.~\ref{Fig_Inc_20}(a) and (b), for which the polarization of
the scattered light is not recorded, one observes that there are
pronounced enhanced backscattering peaks located around the
backscattering direction (at $\theta_s=20^\circ$ and
$\phi_s=180^\circ$). It is also observed that the $p$-polarized
incident beam tends to scatter more light into the forward plane
($q_1>0$) than does an $s$-polarized incident beam.

The first thing to notice from Figs.~\ref{Fig_Inc_20}(c)--(f), where
the polarization of the scattered light is recorded, is that the
co-polarized scattering shows up as an elongated structure with the
long axis of the pattern directed along the plane of incidence, while
the cross-polarized scattering has the long axis of the scattering
pattern perpendicular to this plane. This observation is in agreement
with what was already observed above for normal incidence. However,
for non-normal incidence, the patterns do show less symmetry, as
expected, and an even richer and more complicated angular structure.
In principle, the enhanced backscattering peak phenomenon should exist
in both co- and cross-polarized scattering~\cite{EBP-2,EBP-3,Elena}.
However, for the roughness parameters assumed in this work, one
observes instead of a well-pronounced peak in the backscattering
direction, a ridge of {\em constant} enhanced intensity in parts of
the backscattering plane ($q_1<0$) forming (what seems to be) a half
circle of constant polar scattering angle
$\theta_s\approx\theta_0=20^\circ$ with
$\phi_s\in[90^\circ,270^\circ]$ [Figs.~\ref{Fig_Inc_20}(d) and (e)].
In exactly the backscattering direction, $\theta_s=\theta_0$ and
$\phi_s=180^\circ$, there seems to be little, if any, ``extra''
enhancement in the cross-polarized scattering as compared to the
intensities at other values of $\phi_s$ in the interval
$[90^\circ,270^\circ]$.  The enhancement ridge seen is
Figs.~\ref{Fig_Inc_20}(d) and (e) we speculate is caused by a
constructive interference effect similar in nature to the underlying
enhanced backscattering.

In passing, we note that having available only the in-plane and
out-of-plane results for the same angle of incidence, the local
enhancements observed in {\it e.g.}  Figs.~\ref{Fig1}(b) and
\ref{Fig2}(b) for $\theta_0=20^\circ$, could easily have been mistaken
for well-localized features in the backscattering direction, similar
to what one has for co-polarized scattering [Figs.~\ref{Fig_Inc_20}(c)
and (f)].  In this respect, the angular intensity distributions of the
kind presented in Figs.~\ref{Fig_Inc_00}--\ref{Fig_Inc_40} can provide
important contributions to a better understanding of the 
multiple scattering phenomena. 

Figures~\ref{Fig_Inc_40} present contour plots of the angular
distributions of the mean differential reflection coefficient for a
polar angle of incidence $\theta_0=40^\circ$. Since these results
rather closely resembles those of Figs.~\ref{Fig_Inc_20}, we will not
discuss them further.  However, we note that the structures due to
coherent interference seen in the cross-polarized components for
$\theta_0=20^\circ$, are much harder to identify in the results for
$\theta_0=40^\circ$. This is believed to be caused by the relatively
large angle of incidence, for which it is known that coherent effects
become weaker~\cite{EBP-1}.

\section{Numerical Aspects}
\label{Sec:Numerics}

\begin{table}[tbh]
\begin{center}
\begin{tabular}{ccrrrrrcc}
\hline
\hline
$N$  &  $t_{tot}\mbox{[s]}$  &  $t_A\mbox{[s]}$  & \multicolumn{3}{c}{$t_{A{\bf x}={\bf b}}\mbox{[s]}$} &
$t_{\mathcal E}\mbox{[s]}$ & ${\cal N}$ & ${\cal M}_A\mbox{[Gb]}$ \\
\cline{4-6} 
     &        &         & BiCGStab &\qquad & LU &       &          \\
\hline
64   &  10.5  &  4.0    &   3.5  &&  127  &  3.0  & \;\;8192  & 0.50 \\
80   &  22.0  &  9.5    &   8.0  &&  474  &  4.5  &  12800  & 1.22   \\
100  &  58.5  & 23.0    &  28.5  && 1780  &  7.0  &  20000  & 2.98   \\
112  &  76.0  & 36.0    &  31.0  && 3540  &  9.0  &  25088  & 4.69   \\
\hline
\hline
\end{tabular}
\end{center}
\caption{\label{tab:1} The CPU time spent on various stages of the
  calculations for one realization of the surface profile function and
  one angle of incidence. All CPU times are measured in seconds, and
  the numbers have been rounded to the closest half second, and they
  refer to a machine running an Intel Core2 CPU (Q9550) operating at $2.83\,\mbox{GHz}$
  and running the Linux operating system. The surface was discretized
  on a $N\times N$ grid of points. The reported CPU times are: the
  total CPU time spent for simulating one surface realization for one
  angle of incidence including reading and writing of data
  ($t_{tot}$); the setup of the system matrix of the linear system
  $A{\bf x}={\bf b}$ determining the surface currents ($t_{A}$); the
  time to solve this system by a the iterative BiCGStab method or the
  direct LU decomposition method $(t_{A{\bf x}={\bf b}})$; and finally
  the time to calculate the reflection amplitudes, ${\mathcal E}({\bf
    q}_+,\w)$ for both scattered polarizations on a grid of $101\times 101$
  points ($t_{\mathcal E}$). The number of unknowns to be solved for is  ${\cal
    N}=2N^2$, where the memory (in Gigabytes (Gb)) required to hold the complex system matrix
  $A$, using single precision, is denoted by ${\cal M}_A\propto {\cal N}^2=4N^4$.}  
\end{table}

The rigorous computer simulation approach presented in this work is
rather computationally demanding. Therefore, it is important to be
able to perform such simulations in an efficient manner.  One of the
most challenging aspects of implementing a surface integral method for
a two-dimensional rough surface, is the memory requirement.  By
discretizing the relevant integral equations, in this case
Eq.~(\ref{eq:2}), they are converted into a linear system $A{\bf
  x}={\bf b}$, where $A$ denotes a {\em dense complex} system matrix;
${\bf b}$ is the right-hand-side given in terms of the incident field;
and the unknown vector to be solved for, ${\bf x}$, consists (in our
case) of the independent components of the surface current ${\bf
  J}_H(\bxp |\w)$.  If the randomly rough surface $x_3=\zeta(\bxp)$ is
discretized into $N\times N$ points, then the number of unknowns would
be ${\cal N}=2N^2$, since for a perfectly conducting rough surface we
have {\em two unknowns per surface point} (the two independent
components of ${\bf J}_H$). Hence, the amount of memory needed to hold
the (full) system matrix of the scattering from a perfectly conducting
surface is ${\cal M}_A = 4\,N^4\,m$, where $m$ is the size of a single
scalar complex variable, which on most systems for single and double
precision, respectively, is $m_{S}=8\,\mbox{bytes}$ and
$m_{D}=16\,\mbox{bytes}$.

For each surface realization, there are essentially three
time-consuming steps in this kind of simulation.  They are: ({\it i})
to set up the system matrix; ({\it ii}) to solve the linear system for
the unknown surface currents; and ({\it iii}) to calculate the
reflection amplitudes. Of the three, it is primarily the first two
that are critical and, if not handled properly, particularly the
second. For instance, the total CPU time taken to complete the
calculation using single precision and an iterative solver for one
angle of incidence and one surface realization with $N=112$, including
reading input and writing output data, is $t_{tot}=76.0\,\mbox{s}$ on
an Intel Core2 CPU (Q9550) operating at $2.83\,\mbox{GHz}$ and running
the Linux operating system. On the other hand, for the same simulation
the three steps mentioned above take $t_A=36.0\,\mbox{s}$ to set up
the system matrix, $t_{Ax=b}=31.1\,\mbox{s}$ to solve the linear
system by the use of the iterative BiCGStab method, and $t_{{\cal
    E}}=8.9\,\mbox{s}$ to calculate the reflection amplitudes on a
$101\times101$ grid, in total $76.0\,\mbox{s}$.  Hence, the additional
steps of the calculation, like generating the surface, reading and
writing data to file {\it etc.}, contribute only insignificantly to
the overall CPU time ($t\sim 0.05\mbox{s}$).  The computation times
for other surface discretizations are summarized in Table~\ref{tab:1}.
The reason that it takes a relatively long time (compared to
$t_{tot}$) to set up the matrix elements is the cost of calculating
the exponential function contained in the Green's function.

However, the most critical point to address when trying to reduce the
overall CPU time, is the method used to solve the linear system. In
this work, an iterative solver known as the stabilized bi-conjugated
gradient method (BiCGStab)~\cite{10} has been used, and found to
perform well and to produce reliable results for our application.  The
iteration process of the BiCGStab solver (using a Jacobi
preconditioner) was terminated when the relative error was $10^{-5}$
(or less), which for normal incidence and with $N=112$ required
typically a little more then $20$ iterations when starting from an
initial guess ${\bf x}_{guess}=0$ (of course, other surface parameters
and initial guesses may require more or fewer iterations in order to
reach the desired accuracy).  Using a direct solver, like the
LU-decomposition, would have taken significantly longer (see
Table~\ref{tab:1}). For instance, the time taken to solve the linear
system for $N=112$ by a direct LU solver is $114$ times longer than
that taken by the BiCGStab solver (Table~\ref{tab:1}). Moreover, this
difference is expected to increase with increasing $N$ due to the
different scaling with the number of unknowns (as also shown by the
times presented in Table~\ref{tab:1}). It should be noted that a
direct solver, like the LU-decomposition, opens the possibility for 
carrying out calculations for several angles of incidence (the right-hand sides of the
system) simultaneously with little addition to the overall computation
time. This is not the case for the BiCGStab-method, where the solution
time for several angles of incidence scales linearly with the number
of angles of incidence. There are, however, other iterative methods
that can solve a linear system with several right-hand-sides
with only moderate increase in computational times. One such method is
the (restarted) Generalized Minimal Residual Method~(GMRES)
method~\cite{GMRES}. Compared to the BiCGStab used here, the GMRES is
typically more memory demanding and, therefore, this possibility has
not been explored in this work.

For the sake of comparison, we have repeated the calculations reported
by Tran and Maradudin in Ref.~\cite{1} using the same numerical
parameters (the surface roughness parameters were already the same).
For the calculations carried out in Ref.~\cite{1} solving the integral
equations on a grid of $64 \times 64$ surface points, each iteration
(of which there were six) required $365$ CPU seconds (on a Cray
XMP/EA-116 machine), and to calculate the scattered fields, in-plane
or out-of-plane, required $360$ CPU seconds for each realization of
the surface profile function, for a total of $2550$ CPU seconds for
each realization of the surface profile function.  A similar
calculation required only $7.6$ CPU seconds per surface realization, a
dramatic improvement in speed~\footnote{This time is lower than that
  reported in Table~\ref{tab:1} since only the scattered field in
  either the in-plane or out-of-plane configuration was calculated.}.  This
dramatic reduction occurred for two reasons: First, we have the
overall improvement in general computer hardware. Second, we
hold the whole system matrix in memory due to sufficient memory, while
the approach used in Ref.~\cite{1} was to regenerate the matrix
elements as they were needed. This time cost of the latter is not
insignificant, as we can see from Table~\ref{tab:1}, and both factors
contribute to the overall speedup.

\section{Conclusions}
\label{Sec:Conclusions}

In conclusion, we have shown that the use of the method of moments and
the biconjugate gradient stabilized method provides a formally exact
solution to the problem of the scattering of an electromagnetic field
from a two-dimensional, randomly rough, perfectly conducting surface,
with a modest expenditure of computational time.

Moreover, the full angular distribution of the intensity of the
scattered light, both co- and cross-polarized, was obtained by a
formally rigorous approach for a strongly rough surface. Such
distributions can display rather complex angular patterns that are
rooted in the multiple scattering processes taking place when light
interacts with a strongly rough surface.

Due to the full angular intensity distribution being accessible, the
conservation of energy was checked explicitly for all the calculations
reported and found to be satisfied with an error smaller than $0.5\%$, or
better, something that testifies to the accuracy of the approach and a
satisfactory discretization.


\begin{acknowledgements}
  This research was supported in part by AFRL contract
  FA9453-08-C-0230.  The research of I.S. was in addition supported in
  part by the Research Council of Norway (Sm{\aa}forsk grant) and an
  NTNU Mobility Fellowship.
\end{acknowledgements}

\end{document}